\documentstyle[prl,aps,preprint,tighten]{revtex}
\begin{document}
\draft

        \title{The Upper Critical Field for Unconventional
                Superconducting Film: a Kink due to the Boundary Conditions}
\author{Yu.S.Barash, A.V.Galaktionov, and A.A.Svidzinsky \\  }
\address {I.E. Tamm Department of Theoretical Physics, P.N. Lebedev Physics
Institute \\
     Leninsky Prospect 53, Moscow 117924, Russia \\ }
%\date*{}
\maketitle
\begin{abstract}
Boundary conditions for unconventional superconducting order parameter are
obtained on the basis of a microscopic theory. The upper critical field in a
superconducting film is examined for unconventional superconductors  with
two-component order parameter and is compared with the case of accidental
degeneracy.
It is shown, that for both cases temperature dependences of the upper critical
field in a superconducting film may depend substantially on the quality of
boundaries, and under certain conditions have a kink due to the influence of
film boundaries. The location of a kink point appears to be dependent on a
film thickness. If for a massive sample there is another reason for the
existance of a kink, the interplay of the reasons may lead to a specific
behaviour of a kink point location as a function of a temperature or a film
thickness.
A new test is suggested permitting to distinguish between $E_{1}$ and $E_{2}$
types of pairing in a hexagonal superconductor near $T_c$.

 \end{abstract}
 \pacs{PACS numbers: 74.20.De, 74.60.Ec, 74.50.+r, 74.70.Tx}
 \narrowtext
\section{Introduction}

The properties of unconventional superconductors are currently widely
discussed.
In particular, the types of pairing which correspond to two-component
superconducting order parameter $\bbox{\eta}=(\eta_1,\eta_2)$ are of interest
especially due to the unconventional properties of some heavy-fermion
superconducting  compounds exemplified by $UPt_3$ . Another
possibility of treating the properties of these compounds is related with the
accidental degeneracy, when two different types of pairing  (with
one-component order parameters $\eta_1$ and $\eta_2$ correspondingly) have for
some reasons very close critical temperatures $T_{c1}, T_{c2}$ (see, for
example, \cite{sigueda,sauls,jmvzh,cgarg}). Splitting of the superconducting
phase transition \cite{fish,hass}, a kink (a change in slope) in $H_{c2}(T)$
and
$H_{c1}(T)$ \cite{aden,lin,shiv,vin}, several phase transitions within a mixed
state \cite{aden,mul,qian,schen,klei,bruls,bul,dijk} or under the pressure
\cite{hinks} are the important
experimental evidences in favour of unconventional pairing in $UPt_3$. They
can be described to some extent within the framework of several alternative
approaches
\cite{sigueda,sauls,hts,machoz,ths,bame,machida,mel,zl,zhit,bm,gargc}.
Though the existence of a tetracritical point
on  H-T -phase diagram of superconducting $UPt_3$ at any orientation of a
magnetic field leads to failure if one bases on a theory with two-component
order parameter, the possibility of two neighbouring critical points
(instead of one tetracritical point) is compatible with the theory
\cite{sauls,aden,lin,bruls,dijk,hinks}. So, for the unambiguous identification
of the
type of pairing in $UPt_3$ additional experiments are needed.

In this paper we examine a temperature dependence of the upper critical field
in a thin superconducting film, when hexagonal or tetragonal superconductor is
supposed to be
unconventional with two-component order parameter or with accidentally nearly
degenerate order parameters. It will be shown that the influence of film
boundaries themselves may result in appearance of a kink in $H_{c2}(T)$ at a
temperature $T=T_k$ and the conditions under which it occurs will be
determined.
The value of $T_k$ depends upon a film thickness. For $T_{c1}\ne T_{c2}$ the
kink actually takes place in a massive homogeneous sample as well. The
behaviour
of the kink in a film in comparison with the one in a homogeneous sample has
some distinctive features. Massive superconducting samples of $UPt_3$ appear in
fact to be nonhomogeneous, containing a macroscopic incommensurate structural
modulation \cite{mid}. Considering a thin film as the simplest example for
examining the influence of a nonhomogeneity upon $H_{c2}(T)$, one may come to
the
conclusion about a possibility of a noticeable contribution of lattice
superstructures in $UPt_3$ to the observable kink
of $H_{c2}(T)$. The presence or absense in experiment of the determined below
temperature dependence of $H_{c2}$ for thin superconducting film of $UPt_3$ (or
of any other superconductor) could help to identify the type of pairing
in the sample.

For a solution of the problem it is nessesary to make use of the boundary
conditions for the order parameter in unconventional superconductors. This
problem is examined in the second section of the paper. For this purpose
the microscopic approach is used, based on the quasiclassical Eilenberger
equations.

\section{Boundary Conditions for the Order Parameter in Hexagonal
Unconventional
Superconductors}

The Ginzburg-Landau free energy for a hexagonal unconventional superconductor
with strong spin-orbit coupling and with two-component order parameter
$(\eta_1,\eta_2)$ can be written up to the second order
invariants as follows \cite{vol,sigueda}:
\begin{equation} F=\int dV (a\eta_i\eta_i^* +
K_1p_i^*\eta_j^*p_i\eta_j+
K_2p_i^*\eta_i^*p_j\eta_j+
K_3p_i^*\eta_j^*p_j\eta_i+
K_4p_z^*\eta_i^*p_z\eta_i).
\end{equation}
Here $a=\alpha (T-T_c)$, $p_i=-i\hbar\partial_{ x_i}-(2e/c)A_i$. The $z$
axis is directed along the hexagonal crystalline axis. The indices $i,j$ are
equal to 1 and 2 and correspond to the $x$ and $y$ components for the operator
$\bbox{p}$.We let $\hbar=1$ in this Section.

In the case of one-component order parameter the corresponding free energy may
be obtained from (1) by setting $\eta_2=0$ and $K_2=K_3=0$.

The Ginzburg-Landau equations stemming from (1) should be supplemented by the
boundary conditions for the order parameter. These boundary conditions must be
consistent with the absence  of the current normal to the boundary at the
boundary with vacuum. Since $\bbox{j}=-c\delta F/\delta \bbox{A}$, the
expression
for the superconducting current has the structure:
\begin{equation}
\bbox{j}=-2ie(\eta_1^*\bbox{h}_1+\eta_2^*\bbox{h}_2 -c.c),
\end{equation}
where $\bbox{h}_1,\bbox{h}_2$ are linear combinations of $\partial_i\eta_j$.
The
condition of the vanishing of the current normal to the boundary can be written
as:
\begin{equation}
\left\{ \begin{array}{l}
h_{1n}=b_{11}\eta_1+b_{12}\eta_2\\
h_{2n}=b_{21}\eta_1+b_{22}\eta_2  \end{array} \right.\label{bc} \end{equation}
The quantities in (3) are taken at the boundary, subscript $n$ denotes the
projection on the normal to the boundary $\bbox{n}$ which is directed inwards
for
 definiteness, $b_{11}, b_{22}, b_{12}=b_{21}$ are real. The boundary
conditions
(\ref{bc}) are
 equivalent to ascribing of the effective surface energy to (1) with
density $b_{ij}\eta_i^*\eta_j$. Within the Ginzburg-Landau approach
 $b_{ij}$ are independent upon temperature and they are functions of
 relative orientation of the crystalline axes and
$\bbox{n}$. The values of $b_{ij}$ must be derived from
 microscopic equations. This can be done without loss of generality,
for instance, in the absence of the
 external magnetic field and for the order parameter varying only in the
 perpendicular to the plane boundary direction.
It permits to decouple two equations for the two-component order parameter,
which substantially simplifies the consideration.
Let's denote by $l$ the coordinate along perpendicular to the boundary
direction
and take $l>0$ for definiteness. The generalization on
 the case of the presence of magnetic field and of spatial
 variation of the order parameter parallel to the boundary is
 obvious due to (3).  Generally speaking, in the absence of the external
 magnetic field the current  parallel to the boundary may flow,    so the
 magnetic field may be induced.  Naturally near $T_c$ the account of the
 internal magnetic field  results in corrections to the boundary conditions
 which are small in the measure of $(T_c-T)/T_c$. So such effects usually may
be
neglected and we don't take them into account. Below we shall also consider for
simplicity the case of the spherical Fermi-surface.

When the pairing is singlet, Eilenberger
equations for anisotropic superconductors ( see e.g. \cite{eil,rai1}) can be
reduced to the following system:
  \begin{equation} \left\{ \begin{array}{l}
(2\omega_{m}+ \bbox{v}_{F} \nabla_{\bbox{R}}) f(\hat{\bbox{p}} , \bbox{R},
\omega_{m}) - 2\Delta (\hat{\bbox{p}}, \bbox{R}) g(\hat{\bbox{p}} , \bbox{R},
\omega_{m})=0\\ (2\omega_{m}- \bbox{v}_{F} \nabla_{\bbox{R}})
f^{+}(\hat{\bbox{p}} , \bbox{R}, \omega_{m}) - 2\Delta^{*} (\hat{\bbox{p}},
\bbox{R}) g(\hat{\bbox{p}} , \bbox{R}, \omega_{m})=0\\
 \bbox{v}_{F} \nabla_{\bbox{R}} g(\hat{\bbox{p}} , \bbox{R}, \omega_{m}) +
 \Delta (\hat{\bbox{p}}, \bbox{R}) f^{+}(\hat{\bbox{p}} , \bbox{R}, \omega_{m})
 -\Delta^{*} (\hat{\bbox{p}}, \bbox{R}) f(\hat{\bbox{p}} , \bbox{R},
\omega_{m}) =0
 \end{array}
 \right. \label{eil}\end{equation}
 Here $\omega_{m}=(2m+1)\pi T$ is the Matsubara frequency, $\hat{\bbox{p}}=
 \bbox{p}/|p|$ is the unit vector in the direction of Fermi momentum,
 $\bbox{v}_{F}(\hat{\bbox{p}})$ is the Fermi velocity,
 $\Delta (\hat{\bbox{p}}, \bbox{R})$ is the gap function.  Anomalous Green
functions
 $ f(\hat{\bbox{p}} , \bbox{R}, \omega_{m})$ and $ f^{+}(\hat{\bbox{p}} ,
 \bbox{R}, \omega_{m})= f^{*}(-\hat{\bbox{p}} , \bbox{R}, \omega_{m})$
 and Green function $ g(\hat{\bbox{p}}
 , \bbox{R}, \omega_{m})= g^{*} (-\hat{\bbox{p}} , \bbox{R}, \omega_{m}) $
satisfy
 the normalization condition \begin{equation} g^{2}(\hat{\bbox{p}} , \bbox{R},
 \omega_{m})+ f(\hat{\bbox{p}} , \bbox{R}, \omega_{m}) f^{+}(\hat{\bbox{p}} ,
\bbox{R},
\omega_{m})= 1. \end{equation} The equation of self-consistency is:
\begin{equation} \Delta (\hat{\bbox{p}}, \bbox{R})=-\pi T\sum_{m} \int
\frac{d^{2}S'}{(2\pi)^{3} v_{F}} V(\hat{\bbox{p}},\hat{\bbox{p'}})f(
\hat{\bbox{p'}} ,
 \bbox{R}, \omega_{m}) , \end{equation} where $V(\hat{\bbox{p}},
\hat{\bbox{p'}})$ is
 the anisotropic pairing potential and the integration is carried out over the
 Fermi surface. The system (\ref{eil}) should be supplemented by boundary
conditions
 for quasiclassical propagators \cite{zai,rai3,rai2}. If the nonmagnetic
 impenetrable boundary is specularly reflecting, these boundary
 conditions take the form:
\begin{equation} g(\hat{\bbox{p}})=g(\check{\bbox{p}}), \phantom{d}
f(\hat{\bbox{p}})=f(\check{\bbox{p}}), \phantom{d} f^{+}(\hat{\bbox{p}})=
f^{+}(\check{\bbox{p}}).   \end{equation}
 Here $\hat{\bbox{p}}$ and $\check{\bbox{p}}$ denote the incident and reflected
 momenta correspondingly. According to the mentioned above, expressions for
$b_{ij}$ may be derived without loss of generality in the case
 of the real gap function $\Delta(\hat{\bbox{p}}, \bbox{R})$. Then it is
natural
to
 introduce $f_1=(f+f^+)/2,\: f_2=(f-f^+)/2$. Neglecting higher powers of
$\Delta$ near $T_c$, we come from (4) to the following equation:
\begin{equation}
f_{1}(\hat{\bbox{p}},l, \omega_{m}) -
\frac{v_{l}^{2}}{4\omega_{m}^{2}} \partial_{l}^{2} f_{1}(\hat{\bbox{p}},l,
\omega_{m}) - \frac{\Delta (\hat{\bbox{p}}, l)}{|\omega_{m}|} =0,
\end{equation}
where $v_{l}$ is the $l$-component of the Fermi velocity $\bbox{v}_{F}$.

The solution of this equation satisfying the boundary conditions (7) is:
\begin{equation}
f_{1} (\hat{\bbox{p}},l, \omega_{m})=\frac{1}{|v_{l}|} \int_{0}^{\infty}
\{ \exp(-|\frac{2\omega_{m}}{v_{l}}||l-l'|)\Delta(\hat{\bbox{p}},l')+
\exp(-|\frac{2\omega_{m}}{v_{l}}|(l+l'))\Delta(\check{\bbox{p}},l')\} dl'.
\end{equation}

In the case of one-component order parameter the gap function is supposed to
have a factorized form
 $\Delta (\hat{\bbox{p}}, \bbox{R}) =\eta(\bbox{R}) \psi(\hat{\bbox{p}})$, and
 for the two-component order parameter the corresponding form is
 $\Delta (\hat{\bbox{p}}, \bbox{R}) = \eta_1(\bbox{R}) \psi_1(\hat{\bbox{p}})
+\eta_2(\bbox{R}) \psi_2(\hat{\bbox{p}}) $. Thus the symmetry of the gap
function
in homogeneous superconductor is defined by the symmetry of
$\psi_{1,2} (\hat{\bbox{p}})$.
Substituting (9) into (6) we get with account of these representations two
integral equations in the case of two-component order parameter:

$$\eta_i(l)=\frac{\pi T \lambda}{\int \psi_i^{2}(\hat{\bbox{p}})
d^{2}S}  \sum_{m} \int_{0}^{\infty} \int
d^{2}S  \psi_i (\hat{\bbox{p}})
\{ \frac{\Delta(\hat{\bbox{p}},l')}{|v_{l}|}
\exp(-|\frac{2\omega_{m}}{v_{l}}||l-l'|)+$$
\begin{equation} +\frac{\Delta(\check{\bbox{p}},l')}{
|v_{l}|} \exp(-|\frac{2\omega_{m}}{v_{l}}|(l+l')) \} dl'.
 \end{equation}
Here $\lambda$ is the effective coupling constant in the anisotropic case:
 \begin{equation}
\lambda \psi_i (\hat{\bbox{p}})=-\int V(\hat{\bbox{p}},\hat{\bbox{p'}})
\psi_i(\hat{\bbox{p'}}) \frac{d^{2}S'}{(2\pi)^{3} v_{F}}.
 \end{equation}
It is related to the critical temperature $T_{c}$ by $\pi T_{c}
\lambda \sum_{m} |\omega_{m}|^{-1}=1$. The linearized equations (10) are
applicable at the distances less than $\xi(T)$ from the boundary. The function
$f_2$ does not contribute to them being odd in $\omega_m$.

For the case of the one-component order parameter the equation (10) and
corresponding boundary conditions for $p_x^2-p_y^2$-type of paring in
tetragonal
superconductor have been recently examined in \cite{bgz}.
For those  orientations of the boundary for which the relation
$\psi(\hat{\bbox{p}})=-\psi(\check{\bbox{p}})$ holds, the solution of (10) in
the
case of the one-component order parameter is $\eta(l)=Cl$. Similarly if
the order parameter has the property $\psi(\hat{\bbox{p}})=\psi(
\check{\bbox{p}})$ the
solution of (10) is $\eta(l)=C$. The momenta of electrons reflected from the
boundary are related to the incident momenta by the
reflection accross the plane parallel to the boundary. Since there are seven
planes of symmetry in the symmetry  group $D_{6h}$, we may write the boundary
conditions using the characters of the reflections accross these planes (for
seven
corresponding crystal orientations with
the normal to the boundary perpendicular to the plane of symmetry). If the
character of the one-dimensional irreducible representation is 1 (-1) the
boundary condition is $\eta'(0)=0$ ($\eta(0)=0$). When the integral equations
for two components of the order parameter decouple, the boundary conditions can
be similarly written.

The irreducible representations and corresponding basis functions of $D_{6h}$
are listed in Table I.
We assume that the spins in the triplet state are ordered along $z$ axis, this
is denoted by the spin matrix $\hat{c}$. In this particular case the equations
for the order parameter can be reduced to Eqs.(10) as well.

When the angle $\theta_0$ between the normal to the boundary $\bbox{n}$ and $z$
axis constitutes $\pi/2$ and the azimuthal angle $\phi_0$ (between $x$ and
$\bbox{n}$) equals $k\pi/6$ ($k$ is integer) the boundary conditions for
different representations are enlisted in table II.
The notations
\begin{equation}
\left\{ \begin{array}{l}
\tilde{\eta}_{1}=\eta_1\cos\phi_0+\eta_2\sin\phi_0\\
\tilde{\eta}_{2}=-\eta_1\sin\phi_0+\eta_2\cos\phi_0
\end{array} \right. \label{de1}\end{equation}
are introduced there for $E_{1g}, E_{1u}$ representations.

Similarly for $E_{2g},E_{2u}$ representations:
\begin{equation}
\left\{ \begin{array}{l}
\tilde{\eta}_{1}=\eta_1\cos(2\phi_0)-\eta_2\sin(2\phi_0)\\
\tilde{\eta}_{2}=\eta_1\sin(2\phi_0)+\eta_2\cos(2\phi_0)
\end{array} \right. \label{de2}\end{equation}

If $\bbox{n}$ coincides with $z$ we obtain the following
boundary conditions for one-dimensional representations and for both
components
of two-dimensional representations : for $A_{1u}, A_{2u},\\ B_{1g}, B_{2g},
E_{1g}, E_{2u}$ -- $\eta(0)=0$; for the remaining  representations --
$\eta'(0)=0$. We see that the microscopic structure of the gap function reveals
itself in boundary conditions to the Ginzburg-Landau equations, although these
equations may be similar for different representations.

Now let's discuss the solution of Eq.(10) for arbitrary orientations. It is
easy to check, that for the basis functions of Table I in the case of spherical
Fermi-surface these integral equations decouple, if transformations
(\ref{de1}),(\ref{de2}) are applied for $E_{1},E_{2}$ functions
respectively:
$$\tilde\eta_i(l)=\frac{\pi T \lambda}{\int \tilde\psi_i^{2}(\hat{\bbox{p}})
d^{2}S}  \sum_{m} \int_{0}^{\infty} \int
d^{2}S \tilde\eta_i (l') \tilde\psi_i (\hat{\bbox{p}})
\{ \frac{\tilde\psi_i(\hat{\bbox{p}})}{|v_{l}|}
\exp(-|\frac{2\omega_{m}}{v_{l}}||l-l'|)+$$
\begin{equation} +\frac{\tilde\psi_i(\check{\bbox{p}})}{
|v_{l}|} \exp(-|\frac{2\omega_{m}}{v_{l}}|(l+l')) \} dl'.
\label{fi} \end{equation}
There is no summation over subscript $i$ in this equations and they look
exactly like equations for the one-component order parameter.  The functions
$\tilde \psi_i(\hat{\bbox{p}})$ are related to
$\psi_i(\hat{\bbox{p}})$ by the same equations (\ref{de1}),(\ref{de2}).  The
gap
function may be rewritten as $\Delta(\hat{\bbox{p}}, l)= \tilde\psi_1(
\hat{\bbox{p}}
)\tilde\eta_1(l)+\tilde\psi_2(\hat{\bbox{p}}) \tilde\eta_2(l)$.

At the distances $\xi_0\ll l \ll \xi(T)$ from the boundary Eq.(\ref{fi}) has a
solution $\tilde\eta_i(l)=C(l+q_i)$, the value of $q_i$ may be obtained from
the variational procedure developed in \cite{svi}, which usually turns out to
be a good
approximation.  The exact relation for $q_i$ is:
\begin{equation}
\frac{7\zeta (3)q_i\lambda}{\pi^2 T^2}
\int\tilde\psi_i^{2}(\hat{\bbox{p}})v_l^2d^{2}S=
\frac{\pi\lambda}{48T^3}
\int\tilde\psi_i(\hat{\bbox{p}})(\tilde\psi_i(\hat{\bbox{p}}) +
\tilde\psi_i(\check{\bbox{p}}))|v_l|^3 d^{2}S+[D_{min}\int\tilde
\psi_i^2(\hat{\bbox{p}})d^2S]^{-1},
\end{equation}
where $D_{min}$ is the minimal value of the functional:
\begin{equation}
D=\frac{\int_{0}^{\infty}q(l)[q(l)-\int_{0}^{\infty}K(l,l')q(l')dl']dl}{
[\int_{0}^{\infty}q(l)E(l)]^2}. \label{fu}
\end{equation}
Here $K(l,l')$ is the kernel of the original integral equation (\ref{fi})
$(\tilde\eta_{i}(l)=\int_{0}^{\infty}K(l,l')\tilde\eta_i(l') dl'$)   and
\begin{equation}
E(l)=\pi T\lambda \sum_{m}
\int\tilde\psi_i(\hat{\bbox{p}})\frac{(\tilde\psi_i(\hat{\bbox{p}}) +
\tilde\psi_i(\check{\bbox{p}}))|v_l|}{\omega^2_m}
\exp(-|\frac{2\omega_m}{v_l}|l)d^{2}S.
\end{equation}
The function $q(l)$ in Eq.(\ref{fu}) is supposed to tend to the constant value,
when $l\rightarrow \infty$. Estimating (\ref{fu}) for constant $q(l)$ we obtain
approximate values of $q_i$:
$$
q_i= \biggl(
\frac{\pi^{3}}{336\zeta(3)T}
\int_{v_{l}>0} [\tilde\psi_i(\hat{\bbox{p}})+\tilde\psi_i(\check{\bbox{p}})
]^{2}v_{l}^{3}d^{2}S +  \qquad \qquad \qquad \qquad  \qquad \qquad \qquad
\qquad \qquad \qquad
$$
\begin{equation}
\qquad \qquad \qquad \qquad \qquad
+ \frac{7\zeta(3)}{4\pi^{3}T} \frac{( \int_{v_{l}>0}
[\tilde\psi_i(\hat{\bbox{p}})+\tilde\psi_i(\check{\bbox{p}}
)]^{2}v_{l}^{2}d^{2}S )^{2}}{
\int_{v_{l}>0} [\tilde\psi_i(\hat{\bbox{p}})-\tilde\psi_i(\check{\bbox{p}})
]^{2}v_{l}d^{2}S }\biggr) ( \int\tilde\psi_i^{2}(\hat{\bbox{p}}
)v_{l}^{2}d^{2}S)^{-1}, \label{end}\end{equation}
where the subscript $v_{l}>0$ denotes
that the integration is made over that part of the Fermi surface, where
$v_{l}>0$.  This equation may be applied for one-dimensional representations
directly for the boundary condition $q\eta'(0)=\eta(0)$ with the non-tilded
values $\eta$, since no decoupling problem arises in this case.  Eq.(\ref{end})
remains to be valid for complex basis functions as well after simple
replacement
of squares of the basis functions and of sums and differences of the basis
functions by squares of the corresponding moduli.

Particular values of $q_i$, generally speaking, essentially depend upon the
explicit particular form of basis functions $\psi_i(\hat{\bbox{p}})$, which may
differ even if they belong to the same symmetry representation. Only
values $q=0,\infty$ governed by the symmetry are invariant for the given type
of pairing (irrespective of the shape of the Fermi-surface of the given
symmetry). For the specific basis functions of $E_{1g}$ type from Table I
 the integration in
(\ref{end}) results in:
$$q^s_1=\frac{v_F}{T_c}\sin^22\theta_0\biggl(
%% FOLLOWING LINE CANNOT BE BROKEN BEFORE 80 CHAR
\frac{95\pi^3}{18432\zeta(3)}+\frac{12\zeta(3)}{5\pi^3}\tan^22\theta_0\biggr),$$
\begin{equation}
q^s_2=\frac{v_F}{T_c}\frac{\sin^2\theta_0}{1+2\cos^2\theta_0}\biggl(
\frac{5\pi^3}{1536\zeta(3)}+\frac{4\zeta(3)}{5\pi^3}\tan^2\theta_0\biggr).
\label{e1s}
\end{equation}
The boundary condition takes the form $q_i\tilde\eta_i'(0)=\tilde\eta_i(0)$.
So if  $\theta_0=\pi/2, \phi_0=0 $  we obtain $\eta_1(0)=0$ (since $q_1$
vanishes), while $\eta_2'(0)=0$ (since $q_2=\infty$). Similarly other
orientations
from Table II and the case of $\bbox{n}$ along $z$ may be checked. From (19) it
follows that beyond the
relatively narrow vicinities $\sim (\xi_0/\xi(T))^{1/2}$ of the directions for
which $q_i=\infty$, the parameter $q_i$ has the order of $\xi_0$, because near
these directions
$q_i\sim\xi(0)/(\Delta\theta_0)^2$, where $\Delta\theta_0$ is the angular
displacement from these directions.

For the triplet pairing of $E_1$-type Eq.(\ref{end}) and Table I yield:
$$q_1^t=\frac{v_F}{T_c} \frac{\cos^2\theta_0}{3\sin^2\theta_0+ \cos^2\theta_0}
\biggl(
\frac{5\pi^3}{1344\zeta(3)}+\frac{14\zeta(3)}{15\pi^3}\cot^2\theta_0\biggr),$$
\begin{equation} q_2^t=\infty. \label{e1t} \end{equation} If the functions are
those from Table I of $E_2$ type we obtain in the case of singlet pairing:
$$q_1^s=\frac{v_F}{T_c} \frac{\cos^2\theta_0}{1+2\sin^2\theta_0}
\biggl(
\frac{5\pi^3}{1536\zeta(3)}+\frac{4\zeta(3)}{5\pi^3}\cot^2\theta_0\biggr),$$
\begin{equation} q_2^s=\frac{v_F}{T_c (1+2\sin^2\theta_0)} \biggl(
\frac{5\pi^3}{6144\zeta(3)}(4-4\sin^2\theta_0 +19\sin^4\theta_0)
+\frac{7\zeta(3)}{5\pi^3}\frac{(3\sin^4\theta_0-\sin^2\theta_0+1)^2}{
\sin^22\theta_0}\label{e2s} \biggr).
\end{equation}
And in the case of triplet pairing:
$$q^t_1=\frac{v_F}{T_c}\sin^22\theta_0\biggl(
%% FOLLOWING LINE CANNOT BE BROKEN BEFORE 80 CHAR
\frac{19\pi^3}{4096\zeta(3)}+\frac{32\zeta(3)}{15\pi^3}\tan^22\theta_0\biggr),$$
$$ q_2^t=\frac{v_F\sin^2\theta_0}{T_c
(4+6\sin^2\theta_0-15\sin^4\theta_0+9\sin^6\theta_0)} \biggl(
\frac{\pi^3}{1024\zeta(3)}(100-276\sin^2\theta_0 +195\sin^4\theta_0)+$$
\begin{equation}
+\frac{32\zeta(3)}{15\pi^3}\frac{(22-59\sin^2\theta_0 +41\sin^4\theta_0)^2
\tan^2\theta_0}{
4-12\sin^2\theta_0+27\sin^4\theta_0}\label{e2t} \biggr).  \end{equation}

The values of $q_i$ enable to write down the surface free energy and then
(considering the total free energy) to
 obtain the general boundary conditions for
the Ginzburg-Landau equations, not necessarily for the case, when the order
parameter is changing only in the direction normal to the boundary.
In the weak coupling limit the gradient terms of the
Ginzburg-Landau Hamiltonian, as is known, take the form \cite{wol}:
\begin{equation}
F_{grad}=\frac{7\zeta(3)}{16\pi^2T_c^2}\int
(v_i\partial_i\Delta(\hat{\bbox{p}}
))(v_j\partial_j^*\Delta^*(\hat{\bbox{p}})
)\frac{d^2S}{ (2\pi)^3 v_F}. \end{equation}

So in the case of spherical Fermi-surface and functions from Table I we get,
that for the $E_1$-type of pairing $K_1=K_2=K_3=K$, besides in the case of
singlet pairing $K_4=3K$, while for triplet $K_4=K$. Taking into account these
relations one can find the explicit expressions for $h_{1n}, h_{2n}$ in (3).
Then comparing Eq.(3) with the microscopic boundary conditions
$q_i\tilde\eta_i'(0)
=\tilde\eta_i(0)$ one obtains the coefficients $b_{ij}$ resulting in the
following
expressions for the surface free energy:
$$F^s_{surf}=\frac{3K}{q_1^s(\theta_0)}|\eta_1\cos\phi_0+\eta_2\sin\phi_0|^2+
\frac{K(\sin^2\theta_0+3\cos^2\theta_0)}{
q_2^s(\theta_0)}|\eta_2\cos\phi_0-\eta_1\sin\phi_0|^2, $$ \begin{equation}
F^t_{surf}=\frac{K(3\sin^2\theta_0+\cos^2\theta_0)}{
q_1^t(\theta_0)}|\eta_1\cos\phi_0+\eta_2\sin\phi_0|^2.
\end{equation}
Here $q_i^{s,t}$ are given by (\ref{e1s}),(\ref{e1t}) respectively. Note, that
free energy is written down in terms of original $\eta_1, \eta_2$.

If the pairing is of $E_2$ type, then for the spherical Fermi-surface one has
$K_2=K_3=0$ and $K_4=K/3$ ($K_4=K$) for
 singlet (triplet) pairing. Thus the surface free energy acquires the form:
$$F^s_{surf}=K\biggl(\sin^2\theta_0+\frac{\cos^2\theta_0}{3}\biggr)\biggl\{
\frac{1}{q_1^s(\theta_0)}|\eta_1\cos2\phi_0-\eta_2\sin2\phi_0|^2+
\frac{1}{q_2^s(\theta_0)}|\eta_2\cos2\phi_0+\eta_1\sin2\phi_0|^2\biggr\},$$
\begin{equation}
F^t_{surf}=K\biggl\{
\frac{1}{q_1^t(\theta_0)}|\eta_1\cos2\phi_0-\eta_2\sin2\phi_0|^2+
\frac{1}{q_2^t(\theta_0)}|\eta_2\cos2\phi_0+\eta_1\sin2\phi_0|^2\biggr\}.
\end{equation}

Like the specific values of the parameter $q_i$, the particular form of
the orientational dependence of the surface free energy is governed by the
explicit particular form of the basis functions. Any angular dependence of
$q_i$ and $F_{surf}$ can be certainly expressed in terms of $n_i^2$, i.e.
$\sin^2\theta_0=n_x^2+n_y^2$. The surface free energy can be represented as a
simple quadratic form over $n_x^2, n_y^2, n_z^2$ for a special kind of basis
functions only. According to (24),(25) higher powers of $n_i^2$ are also
involved in the expression for $F_{surf}$ for the basis functions from Table I.
If the free energy (1) is invariant under the rotations on arbitrary angles
around $z$-axis and $q_i$ depend only upon $\theta_0$, one can write down
volume and surface free energies in a $\phi_0$-independent form simultaneously.
It takes place for a particular choice of basis functions for two-dimensional
representations indicated in Table I and for rotationally invariant Fermi
surface. So, in the particular cases (1), (24) or (25) the boundary conditions
in fact do not result in a $\phi_0$-dependence, since $\tilde\eta_i$ can be
used
in the total free energy. One can easily demonstrate that for both
two-dimensional representations there are basis functions breaking rotational
symmetry of the total free energy functional.

If the value of $q_i$ is of the order of $\xi_0$, we have
$\tilde\eta_i(0)\sim \tilde\eta_{i\infty}(\xi_0/\xi(T))\ll \tilde\eta_{i\infty}
$ at the boundary, where $\tilde\eta_{i\infty}$ is the value in the depth of
the superconductor.  So  near $T_c$ we can set approximately $\tilde\eta_i(0)
=0$.
Hence, as it follows from (19), (21), (22), the
prevailing boundary condition (i.e. for the most of orientations of the
boundary) for $E_2$-type of symmetry (both triplet and singlet) and for singlet
pairing of $E_1$- type is:
\begin{equation}
\eta_1(0)=\eta_2(0)=0.
\label{sense}
\end{equation}
For the specific triplet basis functions of $E_1$-type from Table I the
prevailing boundary condition is:
\begin{equation}
\tilde\eta_1(0)=\tilde\eta_2'(0)=0,
\end{equation}
but if other basis functions without rotational symmetry are used, or
non-spherical Fermi-surface is considered, the prevailing boundary condition
will be Eq.(\ref{sense}) for this case as well.

If the scattering at the boundary is diffusive, rather than specular, the
appropriate values of $q_i$ will be of the order of $\xi_0$ as is shown in
\cite{sha}, so we can use Eq.(\ref{sense}) as a good approximation near $T_c$
for all orientations in this case (and for all anisotropic pairings
irrespective
of the crystalline symmetry).

\section{Two-component Order Parameter: $H_{c2}(T) $ for a Superconducting
Film}

Let the crystal axis of high symmetry (axis z) be parallel to the
boundary of a film with thickness $2d$, axis x be perpendicular to the film
surface and a magnetic field $\bbox{H}\parallel x $ (see fig.1).
Though our goal is to describe some properties of a hexagonal superconductor
$UPt_3$  as one of a probable candidate for a
two-component unconventional superconductor, there are
some formal reasons (associated with the simplifying of the problem) to
consider
firstly a more general
case of a tetragonal superconductor. Besides the case of a tetragonal
unconventional superconductor is of interest as itself, since the question
about the type of
pairing remains unsolved also in some tetragonal heavy fermion superconductors
like $CeCu_2Si_2$ and $URu_2Si_2$.
Then the Ginzburg-Landau free energy up to
the second order terms reads $$ F=\int \int dydz
\biggl \{\int\limits_{-d}^{d}dx\left[a\eta_{i}\eta_{i}^{*}+
%% FOLLOWING LINE CANNOT BE BROKEN BEFORE 80 CHAR
K_{1}p_{i}^{*}\eta_{j}^{*}p_{i}\eta_{j}+K_{2}p_{i}^{*}\eta_{i}^{*}p_{j}\eta_{j}+
K_{3}p_{i}^{*}\eta_{j}^{*}p_{j}\eta_{i}
+K_{4}p_{z}^{*}\eta_{i}^{*}p_{z}\eta_{i}+
\right.
$$
\begin{equation}
\left.
+K_{5}(|p_{x}\eta_{2}|^{2}+|p_{y}\eta_{1}|^{2}) \right]
+S_{1}\left(|\eta_{1}|^{2}\mid_{x=-d}+|\eta_{1}|^{2}\mid_{x=d}\right)+
S_{2}\left(|\eta_{2}|^{2}\mid_{x=-d}+|\eta_{2}|^{2}\mid_{x=d}\right)\biggr \}
\label{eqno (3.1)}
\end{equation}

The terms  describing the surface contribution to the free energy in
Eq.(\ref{eqno (3.1)}), correspond to taking into account the boundary
conditions
for the order parameter. We have not written one more surface term
$S_3(\eta_1\eta_2^*+\eta_1^*\eta_2)$ since for the given orientation of crystal
axes relative to the film boundary one has $S_3=0$. The coefficients
$S_1,S_2,S_3$ may be determined by comparing the boundary conditions following
from (\ref{eqno (3.1)}) with those based on a microscopic theory.

It is convenient to choose the vector potential in the form $A_z=Hy$ and to
introduce dimensionless quantities $\varepsilon=(K_2+K_3)/K$,
$\mu=(K_2-K_3)/K$,
 $\zeta=2K_4/K$, $\nu=(K_2+K_3-K_5)/K$, $\tilde a=-ac\zeta^{1/2}/(2\hbar
K_4|e|H)$,
 $\tilde S_{1,2}=2S_{1,2}l/(\hbar^{2}K)$, $x\rightarrow xl$, $y\rightarrow yl$,
$z\rightarrow zl,$ where $K=2K_1+K_2+K_3+K_5$, $l=\zeta^{-1/4}
(\hbar c/2|e|H)^{1/2}$. Then Eq.(\ref{eqno (3.1)}) is transformed as follows
$$
F=\frac{1}{2}K\hbar^{2}l\int dy dz \biggl \{ \int\limits_{-L}^{L} dx [-\tilde
a(|\eta_{1}|^{2}+|\eta_{2}|^{2})+(1+\nu)(|\partial _{x}\eta_{1}|^{2}+|\partial
_{y}\eta_{2}|^{2})+
$$
$$
+(1-\nu)(|\partial
_{x}\eta_{2}|^{2}+|\partial_{y}\eta_{1}|^{2})+y^{2}(|\eta_{1}|^{2}+|
\eta_{2}|^{2})+
(\varepsilon+\mu)(\partial_{x}\eta_{1}^{*}\partial_{y}\eta_{2}+c.c.)+
$$
\begin{equation}
+(\varepsilon-\mu)(\partial_{x}\eta_{2}^{*}\partial_{y}\eta_{1}+c.c.)]+
\tilde S_{1}\left(|\eta_{1}|^{2}|_{x=-L}+|\eta_{1}|^{2}|_{x=L}\right)+
\tilde S_{2}\left(|\eta_{2}|^{2}|_{x=-L}+|\eta_{2}|^{2}|_{x=L}\right)\biggr \}
\label{eqno (3.2)}
\end{equation}
Here $2L=2d/l$ is a dimensionless  film thickness. One should take
into account that for the given gauge of $\bbox{A}$ it is possible to choose
the order parameter $\bbox{\eta}=(\eta_1, \eta_2)$ to be independent of $z$
in considering the upper critical field (it corresponds to the choice of
the origin at the y-axis).

Variation of the free energy functional (\ref{eqno (3.2)}) on
$\eta_1^*, \eta_2^*$ results in the equations
\begin{equation}
\left\{
\begin{array}{l}
\left[-(1+\nu)\partial_{x}^{2}+y^{2}-(1-\nu)\partial_{y}^{2}\right]\eta_{1}-
2\varepsilon\partial_{xy}
\eta_{2}=\tilde a\eta_{1} \\   \\
\left[-(1-\nu)\partial_{x}^{2}+y^{2}-(1+\nu)\partial_{y}^{2}\right]\eta_{2}
-2\varepsilon\partial_{xy}
\eta_{1}=\tilde a\eta_{2}
\end{array}
\right.
\label{3.3}
\end{equation}

and in the boundary conditions
\begin{equation}
\left\{
\begin{array}{l}
\left( (1+\nu)\partial_{x}\eta_{1}+(\varepsilon+\mu)\partial_{y}\eta_{2}\pm
\tilde
S_{1}\eta_{1}\right)_{|x=\pm L}=0 \\
\\
\left( (1-\nu)\partial_{x}\eta_{2}+(\varepsilon-\mu)\partial_{y}\eta_{1}\pm
\tilde S_{2}\eta_{2}
\right)_{|x=\pm L}=0
\end{array}
\right.
\label{3.4}
\end{equation}

Due to linearity of Eqs.(\ref{3.3}, \ref{3.4}) one can consider only those
solutions which are characterized by the definite values of parities relative
to the independent inversions of $x$ and $y$ axes. Then it is easy to see that
the components $\eta_1$ and $\eta_2$ belonging to the same solution of
Eqs.(\ref{3.3}, \ref{3.4}), have opposite values of parities both on $x$ and
on $y$.

The condition of positive definiteness of the sum of the gradient terms
in (\ref{eqno (3.2)}) results in the following restrictions
\begin{equation}
1\pm \nu >|\varepsilon\pm \mu| \ \ \ \ \ \ , \ \  \ \ \ \ \zeta >0
\label{eqno (3.5)}
\end{equation}
In particular, it follows from (\ref{eqno (3.5)}) that  $|\varepsilon|<1$ ,
$|\mu|<1$ ,
$|\nu|<1$. Thus one can usually regard the parameters $\varepsilon, \mu, \nu$
as quite
small (for the majority of their admissible values).We shall employ  below
the perturbation theory  up to the second order in $\varepsilon$ for the
solution of
the problem of the upper critical field for the film. For a hexagonal crystal
one has $K_5=0$ and hence $\nu=\varepsilon$. However, for the simplest
qualitative
examining of the problem it is convenient to assume the parameters $\nu$ and
$\varepsilon$ to be independent and consider only $\varepsilon$ as a small
quantity. It is,
generally speaking, possible for a tetragonal crystal.
 The problem of determining of the minimal eigenvalue $\tilde a$ will be solved
for specular interfaces and for diffusive boundaries separately.

\subsection{Specularly reflecting film surfaces}

Since $yz$-plane is a symmetry plane both for the symmetry group $D_{6h}$ and
for $D_{4h}$, for a given crystal orientations and two-dimensional
representations of those groups specularly reflecting film boundaries
correspond to the limit $\tilde S_1\rightarrow +\infty$, $\tilde S_2=0$ in a
phenomenological Eq.(\ref{3.4}). Then the boundary conditions read
\begin{equation}
\eta_{1|x=\pm L}=0 \ \ \ \ \ \ \ , \ \ \ \ \ \ \ \partial_{x}\eta_{2|x=\pm L}=0
\label{eqno (3.6)}
\end{equation}

The solutions of Eq.(\ref{3.3}) with boundary conditions (\ref{eqno (3.6)}) can
be represented as follows
\begin{equation}
\left\{
\begin{array}{l}
\eta_{1n}=\cos(k_{n}x+\varphi_{n})\eta_{1n}(y)
\\
\ \ \ \ \ \ \ \ \ \ \ \ \ \ \ \ \ \ \ \ \ \ \ \ \ \  \ \ \ \ \ \ \ \  \  \  \ \
\   \ \ \ \ \ \ \ \ \ \ \ \ \ \ \ \ \ \ \ \ \ \ \ \ \ \ \
n=0 \ , \ \pm 1 \ , \ \pm 2 \ , \ \cdots \ \ \ \ \ \
\ \  \\ \eta_{2n}=\sin(k_{n}x+\varphi_{n})\eta_{2n}(y)
\end{array}
\right.
\label{(3.7)}
\end{equation}
where $k_n=(\pi/2L)n$, $\varphi_n=(\pi/2)(n+1)$ and functions $\eta_{1n}(y),
\eta_{2n}(y)$  must satisfy the system of equations
\begin{equation}
\left\{
\begin{array}{l}
(y^{2}-(1-\nu)\partial_{y}^{2})\eta_{1n}(y)-2\varepsilon k_{n}\partial_{y}
\eta_{2n}(y)=
(\tilde a-k_{n}^{2}(1+\nu))\eta_{1n}(y)
\\
\qquad \qquad \qquad \qquad \qquad \qquad \ \ \ \ \ \ \  \ \ \ \ \ \ \ \  \ \ \
\ \ \ \ \ \ \ \ \ \ \ \ \ \ \  \  \ \ \ \ \ \ \ \ \ \ \ \ \ \ \ \ \ \ \ \ \ \ \
\ \ \ n\ne0
\\
(y^{2}-(1+\nu)\partial_{y}^{2})\eta_{2n}(y)+2\varepsilon
k_{n}\partial_{y}\eta_{1n}(y)= (\tilde a-k_{n}^{2}(1-\nu))\eta_{2n}(y)
\end{array}
\right.
\label{(3.8)}
\end{equation}
\begin{equation}
(y^{2}-(1+\nu)\partial_{y}^{2})\eta_{20}(y)=\tilde a\eta_{20}(y)   \qquad ,
\qquad \qquad \qquad n=0
\label{eqno (3.9)}
\end{equation}

The problem of the upper critical field becomes especially simple and
transparent in the zero order approximation in the parameter $\varepsilon$,
when
Eqs.(\ref{(3.8)}) for $\eta_{1n}$, $\eta_{2n}$  decouple. From the right
sides of Eqs.(\ref{(3.8)}) one can see, that dependence of any component of the
order parameter upon $x$ results in decreasing of corresponding ``critical
temperature'' $T_n$ (as compared to $T_c$ in the absence of a field). As a
result all the quantities
$H_n(T)$, following from (\ref{(3.8)}) for $n\ne 0$ as candidates for the upper
critical field, describe the magnetic fields decreasing with increasing
temperature and vanishing at $T_n<T_c$. Furthermore, it follows from the
equations
for $\eta_{2n}$ that the maximum value of magnetic field is reached for the
equation (\ref{eqno (3.9)}) with $n=0$, since $\tilde a, k_n^2\propto 1/H $ and
$\nu<1$. As $T_0=T_c>T_n$ $(n\ne 0)$ the corresponding magnetic field $H_0(T)$
must coincide with $H_{c2}$ close enough to $T_c$. Then one has for the slope
of
the upper
critical field $H^{\prime}_{c2}\propto 1/(1+\nu)^{1/2}$. At the same time it
follows
from the equations for $\eta_{1n}(y)$ that the slopes of corresponding
temperature dependent magnetic fields $\propto 1/(1-\nu )^{1/2}$. Hence for
$\nu >0$ these magnetic fields increase with decreasing of temperature
faster than the field $H_0(T)$, defined by (\ref{eqno (3.9)}). Since the
equation with $n=1$ leads to the maximum admissible value of the magnetic field
among all the equations for $\eta_{1n}$ $(n\ne 0)$, one comes eventually to the
following result at $\varepsilon =0$. In the case $\nu<0$ and for the
specularly
reflecting boundaries the upper critical field for the film  coincides with the
one for the massive sample for all temperatures and is described according to
(\ref{eqno (3.9)}):
\begin{equation}
H_{c2}(T)=\frac{\Phi_{0}}{2\pi\xi^{2}(T)\sqrt{\zeta(1+\nu)}}
\label{eqno (3.10)}
\end{equation}
where $\xi^{2}(T)=\hbar^{2}K/2|a|=\xi_{0}^{2}/(1-T/T_{c})$
, $\Phi_{0}=\pi \hbar c/|e|$ .

But as far the case $\nu>0$  is concerned  the expression (\ref{eqno (3.10)})
describes the upper critical field only within the certain temperature interval
$T_k^{(0)}<T<T_c$, and for $T\le T_k^{(0)}$ the curve $H_{c2}(T)$ suffers a
kink and is described as follows
\begin{equation}
H_{c2}^{(0)}(T)=\frac{\Phi_{0}}{2\pi
\xi_{0}^{2}\sqrt{\zeta(1-\nu)}}\left(1-\frac{T}{T_{c}}-\frac{\pi^{2}
\xi_{0}^{2}}{4d^{2}}(1+\nu)\right)
\label{eqno (3.11)}
\end{equation}
For the point of kink $T_k^{(0)}$ one has from (\ref{eqno (3.10)}),
(\ref{eqno (3.11)})
\begin{equation}
T_{k}^{(0)}=T_{c}\left[1-\frac{\pi^{2}\xi_{0}^{2}}{4d^{2}}\frac{(1+\nu)^{3/2}}
{(\sqrt{1+\nu}-\sqrt{1-\nu})}\right]
\label{eqno (3.12)}
\end{equation}
In order to satisfy the condition $|T_k^{(0)}-T_c|\ll T_c$  the inequality
$d^2\nu\gg \xi_0^2 $ must hold.

Since the solutions with $n=0$ and $n=1$ have a different symmetry--  opposite
parities relative to the inversions on $x$ and on $y$-- for any value of
$\varepsilon$
(and not only for $\varepsilon=0$), it is obvious that the kink at the curve
$H_{c2}(T)$
must remain in the case $\varepsilon \ne 0$ as well ( for the given
orientations
of
magnetic field and crystal axes). In the case $\varepsilon \ne 0$ the
quantitative
consideration of the problem becomes more complicated. However, as for
hexagonal
superconductors $\nu=\varepsilon$, it should be carried out in order to
describe
the
effects in question correctly. Corresponding calculations
are represented in Appendix. According to the solution of Eq.(\ref{(3.8)}) for
$\varepsilon \ne 0$, the slope $H^{\prime}_1(T)$ becomes temperature dependent
even within the
region of applicability of the Ginzburg-Landau theory $T_c-T\ll T_c$. At
$T_k\le T\le T_c$ the slope coincides with the bulk result (\ref{eqno (3.10)}).
At $T\le T_k$ one gets in accordance with (A.25) up to the second order
terms in the parameter $\varepsilon$ inclusively

$$
1-\frac{T}{T_{c}}=\frac{2\pi\zeta^{1/2}\xi_{0}^{2}}{\Phi_{0}}\biggl(
(1-\nu)^{1/2}H+\frac{\pi\Phi_{0}}{8\zeta^{1/2}d^{2}}\biggl(1+\nu-\ \ \ \ \ \ \
\ \ \ \ \ \ \ \ \ \ \ \ \ \ \ \ \ \ \ \ \ \ \ \ \ \ \ \ \ \ \ \ \ \ \ \ \ \ \ \
\ \ \ \ \ \ \ \
$$
\begin{equation}
\ \ \ \ \ \ \ \ \ \ \ \ \ \ \ \ \ \ \ \ \ \ \ \ \ \ \ \ \ \ \ \ \ \ \ \ \
-\frac{\varepsilon^{2}(1-C^{2})^{3/2}}{(1-\nu^{2})^{1/2}}
\int\limits_{0}^{1}\frac{dt}{(1-C^{2}t^{2})^{3/2}}t^{\beta(\nu)-\gamma(\nu,H)}
\biggr)\biggr)
\label{eqno (3.13)}
\end{equation}
where
\begin{equation}
C=\left|\frac{\sqrt{1+\nu}-\sqrt{1-\nu}}{\sqrt{1+\nu}+\sqrt{1-\nu}}\right|
\ , \ \beta=\frac{1}{2}\left(1-\sqrt{\frac{1-\nu}{1+\nu}}\right)  \  ,
\ \gamma(\nu,H)=\frac{\pi\nu\Phi_{0}}{8\zeta^{1/2}d^{2}(1+\nu)^{1/2}H}
\label{eqno (3.14)}
\end{equation}

Under the condition $\varepsilon^2,|\nu|\ll L^2=2\pi d^2\zeta^{1/2}H_1(T)/
\Phi_0 $
the upper critical field below the kink temperature $T_k$ takes the form
\begin{equation}
H_{c2}(T)=\frac{\Phi_{0}}{2\pi\xi_{0}^{2}\zeta^{1/2}\sqrt{1-\nu}}\biggl [
1-\frac{T}{T_{c}}-\frac{\pi^{2}\xi_{0}^{2}}{4d^{2}}\biggl (
1+\nu-\frac{\varepsilon^{2}(1-C^{2})^{3/2}}{(1-\nu^{2})^{1/2}}
\int\limits_{0}^{1}\frac{dy
y^{\beta(\nu)}}{(1-C^{2}y^{2})^{3/2}}\biggr)\biggr]
\label{eqno (3.15)}
\end{equation}
In the limit
$\varepsilon^2/L^2\rightarrow 0$, $\nu/L^2\rightarrow0$ the expression
(\ref{eqno (3.15)}),
becomes exact.

The crossing point of the right lines (\ref{eqno (3.10)}), (\ref{eqno (3.15)})
is
\begin{equation}
\left\{
\begin{array}{l}
T_{k}=T_{c}\left\{1-\frac{\displaystyle\pi^{2}\xi_{0}^{2}}{\displaystyle4d^{2}}
\frac{\displaystyle\sqrt{1+\nu}}{\displaystyle\sqrt
{1+\nu}-\sqrt{1-\nu}}\left(1+\nu-\frac{\displaystyle\varepsilon^{2}(1-C^{2}
)^{3/2}}{\displaystyle(1-\nu^{2})
^{1/2}}\int\limits_{0}^{1}\frac{\displaystyle dy y^{\beta(\nu)}}{
\displaystyle(1-C^{2}y^{2})^{3/2}}
\right)\right\}\\ \\
H_{k}=\frac{\displaystyle\pi\Phi_{0}}{\displaystyle8d^{2}\zeta^{1/2}}
\frac{\displaystyle1}{\displaystyle\sqrt{1+\nu}-\sqrt{1-\nu}}
\left(1+\nu-\frac{\displaystyle\varepsilon^{2}(1-C^{2})^{3/2}}{\displaystyle(1-
\nu^{2})^{1/2}}
\int\limits_{0}^{1}\frac{\displaystyle dy y^{\beta(\nu)}}{\displaystyle (1-
C^{2}y^{2})^{3/2}}\right)
\end{array}
\right.
\label{(3.16)}
\end{equation}

It is assumed in (\ref{(3.16)}) that the kink point $(T_k,H_k)$ lies in the
region where the curve $H_1(T)$ nearly coincides with its asymptotic straight
line. It is valid if $\Phi_0\varepsilon^2/(2\pi d^2\zeta^{1/2})\ll H_k$ which
in fact
is equivalent to the condition $\nu\varepsilon^2\ll 1$.  For $\varepsilon=\nu$
the Eqs. (\ref{eqno (3.10)}), (\ref{eqno (3.15)}),
(\ref{(3.16)}) describe the upper critical field in the film for a hexagonal
unconventional superconductor with two-component order parameter. The
corresponding dependence of $H_{c2}(T)$  is presented in fig.2 (line 1).

Considering the upper critical field for some other orientations of crystalline
axes and then comparing the result with one described just now, one can get
several simple tests for identification of one or another type of pairing.
For instance, one can distinguish between the order parameters transforming
according two-dimensional representations $E_{1g}, E_{1u}$ and $E_{2g},
E_{2u}$,
by examining the temperature dependence of the upper critical
field near $T_c$ in the film with specularly reflecting boundaries for two
different orientations. Since for a hexagonal crystal the x- and y-directions
within the basal plane, generally speaking, are not equivalent, let now axis
y be perpendicular to the film surfaces, axis x be parallel to the boundary
and a magnetic field $\bbox{H}\parallel y$. The results for this particular
case
can be readily obtained from the expressions (\ref{eqno (3.10)}),
(\ref{eqno (3.13)}), if one takes into account that initial Eqs.(\ref{3.3}),
(\ref{eqno (3.6)}) must be transformed to the case along the following way.
In the Ginzburg-Landau equations (\ref{3.3}), written in the crystalline
coordinate system, the only modification to the present case is the
substitution
$y^2\rightarrow x^2$, which corresponds to the field direction along the
y crystalline axis. However, as far as the boundary conditions
(\ref{eqno (3.6)})
are concerned, different modifications should be used for the representations
$E_{1g}, E_{1u}$ and $E_{2g}, E_{2u}$. In the case of the representations
$E_{1g}, E_{1u}$  the order parameter components, according to Eqs.(\ref{de1})
at $\phi_0=\pi/2$, must be interchanged in the boundary conditions
(\ref{eqno (3.6)}) along with $\partial_x \rightarrow \partial_y$. Then the
resulting
equations combined with the boundary conditions exactly coincide with the
initial ones after the formal substitution $x\leftrightarrow y,
\eta_1\leftrightarrow \eta_2 $. Hence the upper critical field remains the same
for both orientations
in the case of $E_{1g}, E_{1u}$- representations, but it is not the case for
the
representations $E_{2g}, E_{2u}$. Indeed, according to Eqs.(\ref{de2})
at $\phi_0=\pi/2$ in the latter case the boundary conditions take the same form
(\ref{eqno (3.6)}) but for the substitution $\partial_x \rightarrow
\partial_y$.
Then
 after the formal substitution $x \rightarrow y$,
$y \rightarrow -x$ the resulting equations  differ from the initial equations
only by the sign of $\varepsilon$ and $\nu$ (for a hexagonal crystal
$\nu=\varepsilon$ ).
Since as was shown above, for specularly reflecting boundaries the kink occurs
only in the case $\nu>0$, we are coming to the assertion that if the order
parameter transforms according to the representations $E_{2g}, E_{2u}$,
the kink must be present for one of the orientations and be absent for the
other. This statement holds also for $\phi_0=(2k+1)\pi/6$, where $k=0,1,2,...$.

For the specific case of a tetragonal superconductor with two-component order
parameter one can similarly suggest the following test permitting to
distinguish
relative signs of the parameters $\varepsilon$ and $\nu$. Indeed, let the
normal
to the boundary  $\bbox{n}$ lie between the crystalline axes x,y  constituting
angle $\pi/4$ with each of them, and $\bbox{H}\parallel \bbox{n}$. The film
boundary coincides in this case with one of the symmetry planes of the group
$D_{4h}$. One can show,
that the Ginzburg-Landau equations and the boundary conditions for this
particular crystal orientation are exactly the same as Eqs.(\ref{3.3}),
(\ref{eqno (3.6)})
after the substitution $\nu \leftrightarrow \varepsilon$, if they are written
for the
functions $\eta_1 \pm \eta_2$ in the variables $(x \pm y)/\sqrt{2}$. Then, if
the parameters $\nu$ and $\varepsilon$ have opposite signs, the kink in the
temperature
dependence of $H_{c2}(T)$ occurs only for one of the orientations. In
the other case, when the parameters are of the same sign the kink is present
(absent) for both orientations simultaneously for $\nu>0$ ($\nu<0$).

Let now temperatures $T_{c1}$ and $T_{c2}$ in coefficients $a_i=\alpha(T-
T_{c,i}) $, $i=1,2$ be not equal to each other and $ \Delta T_c=|T_{c1}-T_{c2}|
\ll T_{c1}$. Then one has instead of  $a(|\eta_1|^2+|\eta_2|^2)$
the terms $a_1|\eta_1|^2+a_2|\eta_2|^2$ in the expression (\ref{eqno (3.1)})
for the free energy. Similar to consideration given above  one
can get that at $T_{c1}<T_{c2}$ and $\nu>0$ the kink of the curve $H_{c2}(T)$
occurs both in a massive homogeneous sample and in a film.
The boundary conditions significantly influence the kink location
in the case $\xi_0^2/d^2\gtrsim \Delta T_c/T_c$. On the other hand
at $T_{c1}>T_{c2}$ and $\nu>0$  the kink doesn't exist in a massive
homogeneous sample but appears in a sufficiently thin film when
$T_1(d)<T_{c2}$. Furthermore in the case $\nu<0$ and $T_{c1}<T_{c2}$
there is no kink of $H_{c2}(T)$ both in a massive homogeneous sample and in a
thin film. At last in the case $\nu<0$ and $T_{c1}>T_{c2}$ there is a kink
in a massive homogeneous sample while it is not present in the case of
sufficiently thin film (at $T_1(d)<T_{c2}$ ).

\subsection{Diffusive electron scattering from the boundaries}

While under the condition of a specular reflection of electrons the
kink of $H_{c2}(T)$ exists only in the case $\nu>0$, it will be shown below
that in the case of diffusively scattering film boundaries the kink  of
$H_{c2}(T)$ always takes place no matter what is a value (and a sign) of the
parameter $\nu$. This assertion can be easily understood qualitatively from
the following simple consideration.

According to the results of Section II, in the case of diffusively scattering
film boundaries the boundary conditions read
\begin{equation}
\eta_{1|x=\pm L}=0 \ \ \ \ \ , \ \ \ \ \ \eta_{2|x=\pm L}=0
\label{eqno (3.17)}
\end{equation}

In zero-order approximation in a small parameter $\varepsilon$ the equations
(\ref{3.3})
decouple and their solutions with account of the boundary conditions
(\ref{eqno (3.17)}) can be represented in the form
$\eta_{i,n}^{(0)}(x,y)=\cos(k_nx+\phi_n)\eta_{i,n}^{(0)}(y)$. Solutions for
$\eta_{2n}^{(0)}(x,y)$ may be obtained from the solutions
$\eta_{1n}^{(0)}(x,y)$
by changing $\nu\rightarrow -\nu$. In contrast to (\ref{eqno (3.6)}) the
boundary
conditions (\ref{eqno (3.17)}) don't allow the value $n=0$ to be admissible
both
for $\eta_{1n}$ and for $\eta_{2n}$. Then the least changing of a critical
temperature due to the boundary conditions is realized for $n=1$ and is
proportional to a corresponding coefficient in front of the term with the
second
order derivative over the coordinate $x$ in Eq.(\ref{3.3}), so that
\begin{equation}
\frac{T_c-T_{1}}{T_c-T_{2}}=\frac{1+\nu}{1-\nu}
\label{3.17'}
\end{equation}

On the other hand one can see from the equations for $\eta_1^{(0)}$ and
$\eta_2^{(0)}$ that a ratio of slope $H^{(0)'}_{1}(T)$ (for $\eta_1^{(0)}$)
to a slope $H^{(0)'}_{2}(T)$ (for $\eta_2^{(0)}$) is
\begin{equation}
\frac{H^{(0)'}_{1}(T)}{H^{(0)'}_{2}(T)}=\biggl(\frac{1+\nu}{1-\nu}\biggr)^{1/2}
\label{3.18}
\end{equation}

Thus at $\nu>0$ the field $H_2(T)$ drops to zero at temperature $T_{2}$ greater
than $T_{1}$ which corresponds in turn to zero value of the field $H_{1}(T_1)$.
However,
since a slope $H^{(0)'}_{1}(T)$  has a greater steepness there appears a
crossing of curves $H^{(0)}_{1}(T)$ and $H^{(0)}_{2}(T)$ which results in the
kink of $H_{c2}(T)$. Under the condition $\nu<0$ one gets from (\ref{3.17'}),
(\ref{3.18}) $T_{2}<T_{1}$ while $|H^{(0)'}_{2}(T)|>|H^{(0)'}_{1}(T)|$. It
leads again to the crossing of the curves $H^{(0)}_{1}(T)$ and
$H^{(0)}_{2}(T)$,
and hence to the appearing of the kink. Thus in zero-order approximation
in the parameter $\varepsilon$ it follows from Eqs. (\ref{(3.8)}),
(\ref{eqno (3.17)})
\begin{equation}
H_{c2}^{(0)}(T)=\frac{\Phi_{0}}{2\pi\xi_{0}^{2}\zeta^{1/2}\left[1+|\nu|sgn
(T-T_{k}^{(0)})\right]^{1/2}}\left[1-\frac{T}{T_{c}}-\frac{\pi^{2}
\xi_{0}^{2}}{4d^{2}}(1-|\nu|sgn (T-T_{k}^{(0)}))\right]
\label{eqno (3.20)}
\end{equation}
\begin{equation}
\left\{
\begin{array}{l}
T_{k}^{(0)}=T_{c}\left[1-
\frac{\displaystyle\pi^{2}\xi_{0}^{2}}{\displaystyle4d^{2}}(2+\sqrt{1-
\nu^{2}})\right]
\\
\\
H_{k}^{(0)}=\frac{\displaystyle\pi
\Phi_{0}}{\displaystyle 8d^{2}\zeta^{1/2}}\left[\sqrt{1+\nu}+\sqrt{1-\nu}
\right]
\end{array}
\right.
\label{(3.21)}
\end{equation}

As it follows from
Appendix (see (A.31)), for $\varepsilon\ne0$ the temperature dependence of the
upper critical field
is described up to the second order terms in $\varepsilon$ inclusively  by the
equation
$$
 1-\frac{T}{T_{c}}=\frac{2\pi\zeta^{1/2}\xi_{0}^{2}}{\Phi_{0}}(1+
\tilde\nu)^{1/2}H
+\frac{\pi^{2}\xi_{0}^{2}}{4d^{2}}\biggl(1-\tilde\nu- \ \ \ \ \ \
\ \ \ \ \ \ \ \ \
\ \ \ \ \ \ \ \ \ \ \ \ \ \ \ \ \ \ \ \ \ \ \ \ \ \ \
$$
\begin{equation}
\ \ \ \ \ \ \ \ \ \ - \frac{4\varepsilon^{2}(1-C^{2})^{3/2}}{\pi^2(1-
\nu^{2})^{1/2}}
\int\limits_{0}^{1}dt\frac{t^{\beta(-\tilde\nu)}}{(1-C^{2}t^{2})^{3/2}}
\sum_{k=1}^{\infty}\frac{k^{2}}{\left(k^{2}-\frac{1}{4}\right)^{2}}
t^{\delta(k,\tilde\nu)} \biggr) ,
\label{eqno (3.22)}
\end{equation}
where $\tilde\nu=|\nu|sgn(T-T_k)$, $\delta(k,\tilde\nu)=\pi\Phi_{0}\sqrt{1
-\tilde\nu}\left(4k^2\frac{1+
\tilde\nu}{1-\tilde\nu}-1\right)/(16\zeta^{1/2}d^{2}H)$.

Thus at $\varepsilon\ne0$ and diffusively scattering
boundaries the upper critical field $H_{c2}(T)$ exhibits a feebly marked
nonlinear
temperature dependence (in measure of the parameter $\varepsilon^2$ usually
multiplied
by a small numerical factor) for any sign of
$\nu$.
Under the condition
$1\ll L^2=2\pi\zeta^{1/2}d^2H/\Phi_0$ one can get the exact asymptotic
behaviour
for the upper critical field
\begin{equation}
H_{c2}(T)=\frac{\Phi_{0}}{2\pi\zeta^{1/2}\xi_{0}^{2}(1+\tilde\nu)^{1/2}}
\left(1-\frac{T}{T_{c}}-\frac{\pi^{2}\xi_{0}^{2}}{4d^{2}}\left(1-\tilde\nu-
\frac{
\varepsilon^{2}(1-C^{2})^{3/2}}{(1-\nu^{2})^{1/2}}\int\limits_{0}^{1}
dt\frac{t^{\beta(-\tilde\nu)}}{(1-C^{2}t^{2})^{3/2}}\right)\right)
\label{eqno (3.23)}
\end{equation}

For $T<T_k$ Eq.(\ref{eqno (3.23)}) completely coincides with the asymptotic
expression (\ref{eqno (3.15)}) obtained for the case of specularly reflecting
boundaries.

In the opposite limit  $1\gg L^2$ one has up to the second order
terms in $\varepsilon$ inclusively (see (A.37))
\begin{equation}
H_{c2}(T)=\frac{\Phi_{0}}{2\pi\zeta^{1/2}\xi_{0}^{2}(1+\tilde\nu)^{1/2}}\frac{
\left(1-\frac{T}{T_{c}}-\frac{\pi^{2}\xi_{0}^{2}}{4d^{2}}\left(1-\tilde\nu
\right)\right)}{\left(1-\frac{\varepsilon^{2}}{(1+\tilde\nu)\tilde\nu}\left(1-
\frac{2\sqrt{1-
\nu^{2}}}{\pi\tilde\nu}\cot\left(\frac{\pi}{2}\sqrt{\frac{1-\tilde\nu}{1+
\tilde\nu}}\right)
\right)\right)}
\label{eqno (3.24)}
\end{equation}

In Eq.(\ref{eqno (3.24)}), which is applicable just below $T_c$, $\tilde\nu$
may be
changed on $\nu$ since it follows from (\ref{(3.21)}) that the kink point lies
at $L\sim1$.

The temperature dependence of $H_{c2}(T)$, described by (49) is presented
in fig.2 (line 2).

If  $T_{c1}\ne T_{c2}$  then, like in the case of specularly reflecting
boundaries, one can easily compare the conditions for the presence or
absence of a kink of $H_{c2}(T)$ in a massive homogeneous sample and in
a thin film with diffusively scattering boundaries. A distinction from the
case of specular reflection appears here at $\nu<0$. Then for $T_{c1}<T_{c2}$
the kink is absent in a massive homogeneous sample, but takes place in a
sufficiently thin film with $T_{1}(d)>T_2(d)$. For $T_{c1}>T_{c2}$
and $\nu<0$ a kink occurs for the both cases.

\section{$H_{c2}(T)$ for a superconducting film in the case of accidental
degeneracy}

 In the case of accidental degeneracy two one-component
superconducting order parameters with different symmetry properties correspond
to critical temperatures $T_{c1}$, $T_{c2}$, which are very close to each other
($|T_{c1}-T_{c2}|=\Delta T_c\ll T_{c1}$).
The Ginzburg-Landau functional for uniaxial superconductor in such a situation
may be represented up to the second order invariants as follows
$$
  F=\int dydz\left\{\int\limits_{-d}^{d}dx\left[a_{1}|\eta_{1}|^{2}
+a_{2}|\eta_{2}|^{2}+K(|p_{x}\eta_{1}|^{2}+|p_{y}\eta_{1}|^{2})+
K^{\prime}(|p_{x}\eta_{2}|^{2}+|p_{y}\eta_{2}|^{2})+
\right.\right.
$$
\begin{equation}
\left. \left.
+K_{4}|p_{z}\eta_{1}|^{2}
+K_{4}^{\prime}|p_{z}\eta_{2}|^{2}\right]
+S_{1}\left(|\eta_{1}|^{2}_{|x=-d}+|\eta_{1}|^{2}_{|x=d}\right)+
S_{2}\left(|\eta_{2}|^{2}_{|x=-d}+|\eta_{2}|^{2}_{|x=d}\right)\right\}
\label{eqno (4.1)}
\end{equation}
Here $a_{1}=\alpha_{1}\left(T-T_{c1}\right)$ ,
$a_{2}=\alpha_{2}\left(T-T_{c2}\right)$. The coherence lengths
$\xi^2_{0\parallel}=\hbar^2K/\alpha_1T_{c1}$,
$\xi^2_{0\perp}=\hbar^2K_4/\alpha_1T_{c1}$,
$\xi'^2_{0\parallel}=\hbar^2K'/\alpha_2T_{c2}$,
$\xi'^2_{0\perp}=\hbar^2K'_4/\alpha_2T_{c2}$ will be used below.

In contrast to (\ref{eqno (3.1)}), where there are gradient terms containing
$\partial_{x}\eta_i\partial_y\eta_j$ with $i\ne j$, the terms of such kind are
absent in (\ref{eqno (4.1)}). In consequence of this fact two Ginzburg-Landau
equations for $\eta_1$ and $\eta_2$ decouple, if the terms of
the fourth order are neglected in the expression for the free energy . This
circumstance essentially simplifies the problem of the upper critical field
in a film. Let's consider the same as in a previous section
orientations of the crystalline axes and of the magnetic field relative to the
film boundaries, and the same gauge for the vector-potential $A_z=Hy$. Let
the boundary conditions for the order parameters $\eta_1$ , $\eta_2$ be
\cite{bc}
\begin{equation}
\eta_{1|x=\pm d}=0 \qquad , \qquad \partial_{x}\eta_{2|x=\pm d}=0
\label{eqno (4.2)}
\end{equation}

According to the results of Sec.II, these boundary conditions are realized for
the given orientations, in
particular, for specularly reflecting boundaries of a hexagonal superconductor,
if the order parameter $\eta_1$ corresponds to one of the one-dimensional
irreducible representations $A_{2g}, A_{2u}, B_{2g}, B_{2u}$ of $D_{6h}$ and
the order parameter $\eta_2$ - to one of the representations $A_{1g}, A_{1u},
B_{1g}, B_{1u} $.

The upper critical field may be represented as $H_{c2}=\max (H_1(T),
H_{2}(T))$,
where the linear functions of temperature $H_1(T)$ and $H_2(T)$
are associated with the solutions of the Ginzburg-Landau equations $\eta_1=0,
\ \  \eta_2=\eta_{20}(y)$ and $\eta_1=\cos(\pi x/2d)\eta_{10}(y), \eta_2=0$
respectively.
Here $\eta_{10}(y)$ and $\eta_{20}(y)$ are related to the zero Landau levels
of corresponding equations, and $H_1(T), H_2(T)$ -- the appropriate magnetic
fields.

Firstly  neglecting the difference between $T_{c1}$ and $T_{c2}$, one has
$T_2=T_c$ and $\Delta T_1=T_c-T_1\propto \xi^2_{0\parallel} $. Then
$H_1^{\prime}(T)/H_2^{\prime}(T)=\xi'_{0\perp}\xi'_{0\parallel}/
 \xi_{0\perp}\xi_{0\parallel}$. Thus the
inclined straight line $H_1(T)$ comes to zero at a lower temperature than
$H_2(T)$ and with decreasing of the temperature crosses the line $H_2(T)$ only
under the condition
$\xi'_{0\parallel}\xi'_{0\perp}> \xi_{0\parallel}\xi_{0\perp}$. If
$T_{c1}<T_{c2}$, the above
condition has the form $\xi'_{0\parallel}\xi'_{0\perp}T_{c2}>
\xi_{0\parallel}\xi_{0\perp}T_{c1}$ and remains for the massive sample as well.
The influence of the film boundaries on the crossing point position becomes
noticeable if $\xi _{0\parallel}^2/d^2\gtrsim \Delta T_c/T_c$.
For $T_{c1}>T_{c2}$ the kink of $H_{c2}(T)$ is absent for a massive sample and
appears for a thin film if $T_{1}(d)<T_{c2}$. Under the condition
$\xi'_{0\parallel}\xi'_{0\perp}T_{c2}< \xi_{0\parallel}\xi_{0\perp}T_{c1}$
 and $T_{c1}>T_{c2}$ the kink occurs for a massive
sample and dissapears with decreasing a film thickness, when the condition
$T_{1}(d)<T_{c2}$ becomes valid.

In the case of diffusively scattering boundaries the boundary conditions are
\begin{equation}
\eta_{1|x=\pm d}=0 \qquad, \qquad \eta_{2|x=\pm d}=0.
\label{(4.3)}
\end{equation}

As just above, let's firstly neglect the difference between $T_{c1}$, $T_{c2}$
and compare linear functions  $H_1(T)$, $H_2(T)$ which are related with the
solutions
$\eta_{1}=\cos\left(\pi x/2d\right)\eta_{10}(y)$ , $\eta_{2}=0$
and  $\eta_{1}=0$ , $\eta_{2}=\cos\left(\pi
x/2d\right)\eta_{20}(y)$ respectively. These solutions satisfy the
Ginzburg-Landau equations
following from (\ref{eqno (4.1)}) together with the boundary conditions
(\ref{(4.3)}). Here $H_{c2}=\max (H_1(T), H_2(T))$. Since for the case
in question one has
\begin{equation}
\frac{\Delta T_{1}}{\Delta
T_{2}}=\frac{\xi^2_{0\parallel}}{\xi'^2_{0\parallel}}
\qquad , \qquad
\frac{H_{1}^{\prime}(T)}{H_{2}^{\prime}(T)}=\frac{\xi'_{0\perp}
\xi'_{0\parallel}}{\xi_{0\perp}\xi_{0\parallel}} ,
\label{eqno(4.4)}
\end{equation}
the condition for $H_{c2}(T)$ to have a kink becomes much more restrictive than
in the case of specular reflection. Indeed, if
$\xi_{0\parallel}>\xi'_{0\parallel}$,
the kink takes
place under the condition $\xi'_{0\perp}/\xi_{0\perp}>
\xi_{0\parallel}/\xi'_{0\parallel}>1$.
In the opposite case $\xi_{0\parallel}<\xi'_{0\parallel}$ the kink occurs if
$\xi'_{0\perp}/\xi_{0\perp}<\xi_{0\parallel}/\xi'_{0\parallel}<1$.
The latter condition means, in particular, that if  the relation
$\xi'_{0\perp}<\xi_{0\perp}$ holds for the coherence lengths in the
perpendicular to the basal plane direction, the opposite
relation $\xi'_{0\parallel}>\xi_{0\parallel}$ must be valid for the coherence
lengths in the parallel direction. Though the relationship
of such kind is, in principle, admissible, it is a quite strict
demand\cite{snos}. It is
in contrast with the result of previous section for two-component order
parameter and the same boundary conditions (\ref{(4.3)}), that the kink in the
temperature dependence of the upper critical field always takes place for a
thin film.

If $T_{c1}\ne T_{c2}$ and, for example, $T_{c1}>T_{c2}$, then the kink occuring
for a massive sample under the condition $\xi'_{0\parallel}\xi'_{0\perp}
T_{c2}<\xi_{0\parallel}\xi_{0\perp}T_{c1}$, remains
at $\xi^{2}_{0\parallel}T_{c1}<\xi'^{2}_{0\parallel}T_{c2}$ for a thin film
with diffusively scattering boundaries too.
In the opposite case $\xi^{2}_{0\parallel}T_{c1}>\xi'^{2}_{0\parallel}T_{c2}$
the kink will disappear with decreasing
of film thickness, when $T_{2}(d)>T_{1}(d)$.

\section{Conclusions}

Boundary conditions for the unconventional superconducting order parameter
have been derived on the basis of microscopic theory for specularly reflecting
boundaries. Using these boundary conditions (and those for diffusive
scattering)
for the problem of the upper
critical field in a superconducting film, the appearance of a kink in a
temperature dependence of $H_{c2}(T)$ has been examined. The presence or
absence of a kink depends upon the ratios of coefficients at gradient terms
in the Ginzburg-Landau functional and upon the quality of boundaries (specular
reflection or diffusive scattering). If due to some
other reasons the kink in a temperature dependence of $H_{c2}(T)$ takes place
even for a massive homogeneous superconducting sample, the study of the
location of a kink point in a thin film from the same compound may give a
useful
information about the type of pairing and the relations between the
Ginzburg-Landau coefficients.

In the case of diffusively scattering boundaries and a two-component
superconducting order parameter the kink turns out to be always present for the
crystalline axes and magnetic field orientations considered above. Then the
location of a kink point is qualitatively defined by the condition
$d\sim \xi(T)$.
At the same time in the case of accidental degeneracy the kink of $H_{c2}(T)$
in a thin film occurs due to the influence of diffusively scattering boundaries
only under quite hard restrictions for the relative values of the coefficients
at gradient terms.

For a two-component order parameter and $\varepsilon \ne0$ the kink, strictly
speaking,
takes place only for considered above orientation of a magnetic field
perpendicular to a crystalline axis of high symmetry. But for small enough
values of $\varepsilon$ and $\mu$ a smooth behaviour of the curve $H_{c2}(T)$
may
exhibit an abrupt change of slope resembling the kink very much.

Recently incommensurate structural macroscopic modulations in heavy fermion
$UPt_3$  have been observed along several crystalline directions \cite{mid}.
Proceeding from the consideration of a thin film as the simplest  model for
studying the influence of an inhomogeneity on the sample properties, it is
natural to suppose that for the appropriate orientations of a magnetic field
smooth inhomogeneities of macroscopic scale may lead to the kink of $H_{c2}(T)$
as well. Note, that in the case of a thin superconducting film and of
two-component order parameter the splitting of the superconducting transition
in
the absence of magnetic field may also be provoked due to the boundary
conditions \cite{sigueda,sigog,ogawa}. So, it is possible to consider the
structural
macroscopic modulations as one of the probable sources for the kink of
$H_{c2}(T)$ and for the splitting of the critical temperature in $UPt_3$.

After this article had already been submitted a paper by K.V.Samokhin appeared
in Zh.Eksp.Teor.Fiz. {\bf 107}, 906 (1995), where the boundary conditions for
unconventional superconductors were also considered and applied to the problem
of the surface critical field $H_{c3}$.

The research described in this publication
was made possible in part by  Grant No M2D000 from the International Science
Foundation.

\section*{Appendix}

It is convenient to look for the minimal eigenvalue $\tilde a$ of the problem
(\ref{3.3}) using the perturbation theory in the parameter $\varepsilon$ and
transforming the problem to the form

$$
\hat H_{0}\Psi+\hat V\Psi=\tilde a\Psi
\eqno (A.1)
$$
where $$ \Psi=
\left(
\begin{array}{l}
\eta_{1}\\
\eta_{2}
\end{array}
\right) ,\qquad \qquad
\hat V=
\left(
\begin{array}{cc}
0 & -2\varepsilon\partial_{xy} \\
-2\varepsilon\partial_{xy} & 0
\end{array}
\right) ,
$$
\vspace{0.4cm}
$$
\hat H_{0}=
\left(
\begin{array}{cc}
-(1+\nu)\partial_{x}^{2}+y^{2}-(1-\nu)\partial_{y}^{2}& 0 \\
0& -(1-\nu)\partial_{x}^{2}+y^{2}-(1+\nu)\partial_{y}^{2}
\end{array}
\right).
\eqno (A.2)
$$

In terms of the operators
$$
\hat A=\frac{i\sqrt{1-\nu}\partial_{y}+iy}{\sqrt{2}(1-\nu)^{1/4}} ,\ \
\hat A^{+}=\frac{i\sqrt{1-\nu}\partial_{y}-iy}{\sqrt{2}(1-\nu)^{1/4}},\ \
\hat B=\frac{i\sqrt{1+\nu}\partial_{y}+iy}{\sqrt{2}(1+\nu)^{1/4}} ,\ \
\hat B^{+}=\frac{i\sqrt{1+\nu}\partial_{y}-iy}{\sqrt{2}(1+\nu)^{1/4}}
\eqno (A.3)
$$
\vspace{0.3cm}
which satisfy the conditions
$[\hat A,\hat A^{+}]=1,  [\hat B,\hat B^{+}]=1$, the unperturbed Hamiltonian
reads
$$
\hat H_{0}=
\left(
\begin{array}{cc}
-(1+\nu)\partial_{x}^{2}+2\sqrt{1-\nu}(\hat A^{+}\hat A+\frac{1}{2})& 0\\
0& -(1-\nu)\partial_{x}^{2}+2\sqrt{1+\nu}(\hat B^{+}\hat B+\frac{1}{2})
\end{array}
\right)
\eqno (A.4)
$$

{}From Eq.(A.4) one can see that a complete orthonormal set of eigenfunctions
of
the operator
$\hat H_0$ may be represented as a set $ \{ \Psi_{mn}\}=\{\Psi_{mn}^A$,
$\Psi_{mn}^B\}$, where
$$
\Psi_{mn}^{A}=
\left(
\begin{array}{l}
g_{m}^{A}f_{n}^{A}\\
0
\end{array}
\right)
\qquad , \qquad
\Psi_{mn}^{B}=
\left(
\begin{array}{l}
0\\
g_{m}^{B}f_{n}^{B}
\end{array}
\right).
\eqno (A.5)
$$
Here $g_m^A(x),g_m^B(x)$ -- the orthonormal eigenfunctions of the operator
$\partial^2_x$:
$$
\partial_{x}^{2}g_{m}^{A}=-\lambda _{m}^{A}g_{m}^{A} \qquad , \qquad
\partial_{x}^{2}g_{m}^{B}=-\lambda _{m}^{B}g_{m}^{B}
\eqno (A.6)
$$
They are supposed to be defined at the segment $[-L,L]$ and to satisfy the
boundary conditions for $\eta_1$, $\eta_2$\ respectively.

The  functions
$f_n^A(y)$, $f_n^B(y)$ are the orthonormal wave functions of harmonic
oscilators satisfying the conditions
$\hat A^{+} \hat Af_{n}^{A}(y)=nf_{n}^{A}(y)$  ,   $\hat B^{+} \hat
Bf_{n}^{B}(y)=nf_{n}^{B}(y) $. They may be represented as
follows
$$
f_{0}^{A}=\frac{1}{\pi^{1/4}(1-\nu)^{1/8}}e^{-\frac{y^{2}}{2
(1-\nu)^{1/2}}} \qquad ,\qquad \hat Af_{0}^{A}=0\qquad ,\qquad
f_{n}^{A}=\frac{1}{(n!)^{1/2}}\left(\hat A^{+}\right)^{n}f_{0}^{A}
\eqno (A.7)
$$

$$
f_{0}^{B}=\frac{1}{\pi^{1/4}(1+\nu)^{1/8}}e^{-\frac{y^{2}}{2
(1+\nu)^{1/2}}} \qquad ,\qquad \hat Bf_{0}^{B}=0\qquad ,\qquad
f_{n}^{B}=\frac{1}{(n!)^{1/2}}\left(\hat B^{+}\right)^{n}f_{0}^{B}
\eqno (A.8)
$$

For the perturbation operator one has $$(\Psi^A_{mn})^+\hat V\Psi^A_{kl}=0,
\qquad (\Psi^B_{mn})^+\hat V\Psi^B_{kl}=0,\qquad (\Psi^B_{mn})^+
\hat V\Psi^A_{kl} =
 -2\varepsilon(g_m^{B*}\partial_x g_k^A)(f_n^{B*}\partial_yf_l^A)
\eqno (A.9)
$$

Due to relations (A.5), (A.9), the expression for the minimal value of
$\tilde a$
is $\tilde a=\min\{E^{A},E^{B}\}$, where  the quantities $E^{A},E^{B}$  are
reduced
in the second order of perturbation theory in the operator $\hat V $ to the
form
$$
E^{A}=E_{00}^{A}-\sum\limits_{m,n}^{}\frac{4\varepsilon^{2}|<g_{m}^{B}|
\partial_{x}|g_{0}^{A}>
|^{2} |<f_{n}^{B}|\partial_{y}|f_{0}^{A}>|^{2}}{E_{mn}^{B}-E_{00}^{A}}
\eqno (A.10)
$$

$$
E^{B}=E_{00}^{B}-\sum\limits_{m,n}^{}\frac{4\varepsilon^{2}|<g_{m}^{A}|
\partial_{x}|g_{0}^{B}>
|^{2} |<f_{n}^{A}|\partial_{y}|f_{0}^{B}>|^{2}}{E_{mn}^{A}-E_{00}^{B}}
\eqno (A.11)
$$

Eigenvalues of the unpertubed Hamiltonian $\hat H_0$ may be written as
$$
\begin{array}{l}
E_{mn}^{A}=(1+\nu)\lambda_{m}^{A}+2\sqrt{1-\nu}(n+\frac{1}{2})\\ \\
E_{mn}^{B}=(1-\nu)\lambda_{m}^{B}+2\sqrt{1+\nu}(n+\frac{1}{2})
\end{array}
\eqno (A.12)
$$

Inserting (A.7), (A.8) into the matrix element
$<f_{n}^{A}|\partial_{y}|f_{0}^{B}>$,
one gets
$$
<f_{n}^{A}|\partial_{y}|f_{0}^{B}>=\int\limits_{-\infty}^{\infty}f_{n}^{A^*}
\partial_{y}f_{0}^{B}
dy=
$$
$$
=\frac{-i^{n}(1-\nu)^{n/4}}{(n!)^{1/2}2^{n/2}\pi^{1/2}
(1-\nu^{2})^{1/8}(1+\nu)^{1/2}}\int\limits_{-\infty}^{\infty}
ye^{-\frac{y^{2}}{2(1+\nu)^{1/2}}}\left(\frac{y}{(1-\nu)^{1/2}}-
\partial_{y}\right)^{n}
e^{-\frac{y^{2}}{2(1-\nu)^{1/2}}}dy
\eqno (A.13)
$$

The relation
$$
I=\int\limits_{-\infty}^{\infty}dye^{-\frac{\beta y^{2}}{2}}y(\alpha
y-\partial_{y})^{m}e^{-\frac{\alpha y^{2}}{2}}=\sqrt{2\pi}2^{\frac{m+1}{2}}
\alpha^{\frac{m+1}{2}}
\frac{(\alpha-\beta)^{\frac{m-1}{2}}}{(\alpha+\beta)^{\frac{m}{2}+1}}m!!
\qquad , \qquad m=1,3\ldots
\eqno (A.14)
$$
permits to transform (A.13) to
$$
<f_{n}^{A}|\partial_{y}|f_{0}^{B}>=
\left\{
\begin{array}{ccc}
0& , & n=0,2,4,\ldots\\
-\frac{\displaystyle i^{n}\sqrt{2}(1-\nu)^{3/8}(1+\nu)^{1/8}(n!!)C^{n/2}
(sgn \nu)^{(n-1)/2}}{\displaystyle(n!)^{1/2}(\sqrt{1+\nu}+\sqrt{1-\nu})
^{1/2}|\nu|^{1/2}} &
, & n=1,3,5,\ldots,
\end{array}
\right.
\eqno (A.15)
$$

\noindent where
$$
C=\left|\frac{\sqrt{1+\nu}-\sqrt{1-\nu}}{\sqrt{1+\nu}+\sqrt{1-\nu}}\right|=
\frac{|\nu|}{1+(1-\nu^{2})^{1/2}}  .
\eqno (A.16)
$$

The expression for the matrix element $<f_{n}^{B}|\partial_{y}|f_{0}^{A}>$ may
be
obtained from (A.15) by changing $\nu \rightarrow -\nu$.

With the aid of (A.12), (A.15) the Eq.(A.11) is reduced to
$$
E^{B}=(1-\nu)\lambda
_{0}^{B}+\sqrt{1+\nu}- \qquad \qquad \qquad \qquad \qquad \qquad \qquad \qquad
\qquad \qquad \qquad \qquad \qquad \qquad
$$
$$
\qquad \qquad \qquad  \qquad -\frac{\varepsilon^{2}(1-C^{2})^{3/2}}{(1-\nu^{2})
^{1/2}}\sum\limits_{
m=0,1,2,\ldots \atop n=1,3,5,\ldots
}^{}\frac{|<g_{m}^{A}|\partial_{x}|g_{0}^{B}>|^{2}C^{n-1}(n!!)^{2}}{n!\left[
n+\frac{1}{2}\left(1-\sqrt{\frac{1+\nu}{1-\nu}}\right)+\frac{1}{2}\sqrt{1-\nu}
\left(\frac{1+\nu}{1-\nu}\lambda_{m}^{A}-\lambda_{0}^{B}\right)\right]}
\eqno (A.17)
$$

The sum over n is calculated along the following way
$$
\sum\limits_{n=1,3,\ldots}^{}\frac{C^{n-1}(n!!)^{2}}{n!(n+\lambda)}=
\sum\limits_{n=1,
3,\ldots}^{}\int\limits_{0}^{1}
\frac{C^{n-1}(n!!)^{2}t^{n+\lambda-1}}{n!}dt=\int\limits_{0}^{1}t^{\lambda}
\sum\limits_{n=1,3,\ldots}^{}\frac{(Ct)^{n-1}(n!!)^{2}}{n!}dt=
$$
$$
=\int\limits_{0}^{1}
\frac{t^{\lambda}}{(1-(Ct)^{2})^{3/2}}dt=\frac{1}{C^{\lambda+1}}
\int\limits_{0}^{C}
\frac{t^{\lambda}}{(1-t^{2})^{3/2}}dt=\frac{1}{\sqrt{1-C^{2}}}-
\frac{\lambda}{C^{\lambda+1}}\int\limits_{0}^{\arcsin C}\sin^{\lambda}(y)dy .
\eqno (A.18)
$$

Introducing the designation
$\beta(\nu)=[1-(1-\nu)^{1/2}/(1+\nu)^{1/2}]/2$ and using (A.18), one can
transform (A.17) to the following expression
$$
E^{B}=\sqrt{1+\nu}+(1-\nu)\lambda_{0}^{B}- \qquad \qquad \qquad \qquad \qquad
\qquad
$$
$$
\qquad \qquad \qquad -\frac{\varepsilon^{2}(1-C^{2})^{3/2}}{(1-\nu^{2})
^{1/2}}\int\limits_{0}^{1}\frac{t^{\beta(-\nu)}dt}{(1-C^{2}t^{2})^{3/2}}
\sum\limits_{m=0,1,2,\ldots}^{}|<g_{m}^{A}|\partial_{x}|g_{0}^{B}>|^{2}
t^{\gamma_{m0}^{AB}(\nu)} \ \ ,
\eqno (A.19)
$$
where $\gamma_{m0}^{AB}(\nu)=\sqrt{1-\nu}\left [\lambda_{m}^{A}(1+\nu)/(1-\nu)-
\lambda_{0}^{B}\right ]/2$.

For $E^A$ one gets in analogous way
$$
%\begin{array}{l}
E^{A}=\sqrt{1-\nu}+(1+\nu)\lambda_{0}^{A}- \qquad \qquad \qquad \qquad \qquad
\qquad
$$
$$
\qquad \qquad \qquad -\frac{\varepsilon^{2}(1-C^{2})^{3/2}}{(1-\nu^{2})
^{1/2}}\int\limits_{0}^{1}\frac{t^{\beta(\nu)}dt}{(1-C^{2}t^{2})^{3/2}}
\sum\limits_{m=0,1,2,\ldots}^{}|<g_{m}^{B}|\partial_{x}|g_{0}^{A}>|^{2}
t^{\gamma_{m0}^{BA}(\nu)},
%\end{array}
\eqno (A.20)
$$
where $ \gamma_{m0}^{BA}(\nu)=\sqrt{1+\nu}\left
[\lambda_{m}^{B}(1-\nu)/(1+\nu)-
\lambda_{0}^{A}\right]/2$.

Since one should use now the explicit expressions for functions $g_{m}^{A}$,
$g_{m}^{B}$, below we consider the cases of specularly reflecting and
duffusively
scattering boundaries separately.

\subsection*{a) Specularly reflecting boundaries}

For the given crystal orientation the boundary conditions for functions
$g_{m}^{A}$, $g_{m}^{B}$ in the case of specularly reflection are
$$
g^{A}_{m|x=\pm L}=0 \qquad , \qquad \partial_{x}g^{B}_{m|x=\pm L}=0
\eqno (A.21)
$$
The orthonormal eigenfunctions of the operator $\partial_x^2$ satisfying the
conditions (A.21),
constitute the sets
$$
g_{m}^{A}=\frac{1}{\sqrt{L}}\cos(k_{m+1}x+\phi_{m+1}) \qquad , \qquad
g_{m}^{B}=\frac{1}{\sqrt{L}}\sin(k_{m}x+\phi_{m})
\eqno (A.22)
$$
where  $k_{m}=\frac{\pi}{2L}m$ \quad , \quad $\phi_{m}=\frac{\pi}{2}(m+1)$
\quad , \quad $m=0,1,2,\ldots $

Substituting (A.22) into (A.19), (A.20) and taking into account relations
$$
<g_{m}^{A}|\partial_{x}|g_{0}^{B}>=0 \qquad , \qquad
<g_{m}^{B}|\partial_{x}|g_{0}^{A}>=
\left\{
\begin{array}{ccc}
 0& , & m\ne 1\\
-k_{1}& , &m=1
\end{array}
\right.
\eqno (A.23)
$$
one obtains
$$
E^{B}=\sqrt{1+\nu}
\eqno (A.24)
$$

$$
%\begin{array}{l}
E^{A}=\sqrt{1-\nu}+\left(\frac{\pi}{2L}\right)^2\left(1+\nu-
\frac{\varepsilon^{2}(1-C^{2})^{3/2}}{(1-\nu^{2})
^{1/2}}\int\limits_{0}^{1}\frac{dtt^{\beta(\nu)}}{(1-C^{2}t^{2})^{3/2}}
t^{\frac{-\nu\pi^2}{4L^2\sqrt{1+\nu}}}\right)
%\end{array}
\eqno (A.25)
$$

Eq.(A.25),which is applicable for small enough $\varepsilon$ and any
admissible value of $\nu$, is reduced in the limit $|\nu|\ll L^2$ to the
following expression
$$
|\nu|\ll L^{2} \quad : \qquad
E^{A}=\sqrt{1-\nu}+\left(\frac{\pi}{2L}\right)^2\left(1+\nu-
\frac{\varepsilon^{2}(1-C^{2})^{3/2}}{(1-\nu^{2})
^{1/2}}\int\limits_{0}^{1}\frac{dtt^{\beta(\nu)}}{(1-C^{2}t^{2})^{3/2}}
\right)
\eqno (A.26)
$$

\subsection*{b) Diffusively scattering boundaries}

Under the boundary conditions
$$
g^{A}_{m|x=\pm L}=0 \qquad , \qquad g^{B}_{m|x=\pm L}=0
\eqno (A.27)
$$
the corresponding eigenfunctions of the operator $\partial^2_x$ may be
written as
$$
g_{m}^{A}=g_{m}^{B}=\frac{1}{\sqrt{L}}\cos(k_{m+1}x+\phi_{m+1}) \qquad , \qquad
\eqno (A.28)
$$
where  $k_{m}=m\pi/2L$  ,  $\phi_{m}=(m+1)\pi/2$
 ,  $m=0,1,2,\ldots $.

{}From (A.28) one gets the following expressions for matrix elements
$$
%% FOLLOWING LINE CANNOT BE BROKEN BEFORE 80 CHAR
<g_{m}^{A}|\partial_{x}|g_{0}^{B}>=\frac{1}{L}\int\limits_{-L}^{L}\cos(k_{m+1}x+
\phi_{m+1})\partial_{x}\cos(k_{1}x+\phi_{1})dx=
$$
$$
%% FOLLOWING LINE CANNOT BE BROKEN BEFORE 80 CHAR
=\frac{2k_{1}}{\pi}\left(1-(-1)^{m}\right)\left(\frac{1}{m+2}+\frac{1}{m}\right)
\qquad  \qquad m=0,1,2,\ldots
\eqno (A.29)
$$
and for the eigenvalues of the operator $\partial^2_x$:
$\lambda_{0}^{B}=k_{1}^{2}=\left(\frac{\pi}{2L}\right)^{2},
\lambda_{m}^{A}=k_{m+1}^{2}=\left(\frac{\pi}{2L}\right)^{2}(m+1)^{2}$.

With the aid of (A.29) the relation (A.19) takes the form
$$
%\begin{array}{l}
E^{B}=\sqrt{1+\nu}+(1-\nu)\left(\frac{\pi}{2L}\right)^{2}-
$$
$$
-\frac{\varepsilon^{2}(1-C^{2})^{3/2}}{(1-\nu^{2})
^{1/2}}\int\limits_{0}^{1}\frac{dtt^{\beta(-\nu)}}{(1-C^{2}t^{2})^{3/2}}
\sum\limits_{k=1}^{\infty}\frac{1}{L^{2}}\frac{k^{2}}{\left(k^{2}-
\frac{1}{4}\right)^{2}}
t^{\left(\frac{\pi}{2L}\right)^{2}\frac{\sqrt{1-\nu}}{2}
\left(\left(\frac{1+\nu}{1-\nu}\right)4k^{2}-1\right)}
%\end{array}
\eqno (A.30)
$$

Since $g_m^A=g_m^B$, the expression for $E^A$ may be obtained from $E^B$ simply
by substitution $\nu \rightarrow -\nu$. It is easy to see that the curves
$E^B(L)$ and $E^A(L)$ cross each other. Designating $\tilde\nu=
|\nu|sgn\left(T-T_{k}\right)$, where $T_k$ is the point of crossing, one
obtains
for $\tilde a=\min\left(E^{A}, E^{B}\right)$:
$$
\tilde a=\sqrt{1+\tilde\nu}+(1-\tilde\nu)\left(\frac{\pi}{2L}\right)^{2}-
\samepage
$$
$$
-\frac{\varepsilon^{2}(1-C^{2})^{3/2}}{L^{2}(1-\nu^{2})
^{1/2}}\int\limits_{0}^{1}\frac{dtt^{\beta(-\tilde\nu)}}{(1-C^{2}t^{2})^{3/2}}
\sum\limits_{k=1}^{\infty}\frac{k^{2}}{\left(k^{2}-\frac{1}{4}\right)^{2}}
t^{\left(\frac{\pi}{2L}\right)^{2}\frac{\sqrt{1-\tilde\nu}}{2}
\left(\left(\frac{1+\tilde\nu}{1-\tilde\nu}\right)4k^{2}-1\right)}
\eqno (A.31)
$$

Eq.(A.31) can be simplified in the limits of large and small values of
parameter
$L$. For $L^2\gg1$ it follows
$$
L^{2}\gg 1 \quad : \qquad
\tilde a=\sqrt{1+\tilde\nu}+\left(\frac{\pi}{2L}\right)^2\left(1-\tilde\nu-
\frac{\varepsilon^{2}(1-C^{2})^{3/2}}{(1-\nu^{2})
^{1/2}}\int\limits_{0}^{1}\frac{dtt^{\beta(-\tilde\nu)}}{(1-C^{2}t^{2})
^{3/2}}\right) ,
\eqno (A.32)
$$
where the relation
$$
\sum_{k=1}^{\infty}\frac{k^{2}}{\left(k^{2}-\frac{1}{4}\right)^{2}}=
\frac{\pi^{2}}{4}
$$
has been taken into account.
In the opposite limit $L^2\ll1$ the integral in (A.31) is transformed as
follows
$$
\frac{1}{L^{2}}\int\limits_{0}^{1}dt\frac{t^{\beta(-\tilde\nu)}}{(1-
C^{2}t^{2})^{3/2}}
t^{\delta(\tilde\nu)}=
\frac{8}{\pi^{2}\sqrt{1-\tilde\nu}\left(\frac{1+\tilde\nu}{1-\tilde\nu}4k^{2}-
1\right)}
\int\limits_{0}^{1}\frac{t^{\beta(-\tilde\nu)}dt}{(1-C^{2}t^{2})^{3/2}}
t^{\delta(\tilde\nu)}=
$$
$$
=\frac{8}{\pi^{2}\sqrt{1-\tilde\nu}\left(\frac{1+\tilde\nu}{1-\tilde\nu}4k^{2}-
1\right)}
\left(\frac{1}{(1-C^{2})^{3/2}}-
\int\limits_{0}^{1}
t^{\delta(\tilde\nu)}
d\left(\frac{t^{\beta(-\tilde\nu)}}{(1-C^{2}t^{2})^{3/2}}\right)\right)
\stackrel{L\rightarrow 0}{\rightarrow}
$$
$$
\qquad \qquad \qquad  \rightarrow \frac{8}{\pi^{2}\sqrt{1-
\tilde\nu}\left(\frac{1+\tilde\nu}{1-\tilde\nu}4k^{2}-1\right)}
\frac{1}{(1-C^{2})^{3/2}}
\eqno (A.33)
$$
Here $$\delta(\tilde\nu)=\left(\frac{\pi}{2L}\right)^{2}\frac{\sqrt{1-
\tilde\nu}}{2}
\left(\left(\frac{1+\tilde\nu}{1-\tilde\nu}\right)4k^{2}-1\right)
\eqno (A.34)
$$
\\
Then one has
$$
L^{2}\ll 1 \  : \ \
\tilde a=\left(\frac{\pi}{2L}\right)^{2}(1-\tilde\nu)+\sqrt{1+\tilde\nu}
\left(1-\frac{2\varepsilon^{2}}{\pi^{2}(1+\tilde\nu)^{2}}
\sum_{k=1}^{\infty}\frac
{k^{2}}{\left(k^{2}-\frac{1}{4}\right)^{2}}\frac{1}{\left(k^{2}-\frac{(1-
\tilde\nu)}{4(1+
\tilde\nu)}\right)}\right)
\eqno (A.35)
$$

The sum presented here is

$$
\sum_{k=1}^{\infty}\frac{k^{2}}{\left(k^{2}-\frac{1}{4}\right)^{2}}
\frac{1}{(k^{2}-\gamma
%% FOLLOWING LINE CANNOT BE BROKEN BEFORE 80 CHAR
^{2})}=\frac{\pi^{2}}{1-4\gamma^{2}}\left(1-\frac{8\gamma\cot(\pi\gamma)}{\pi(1-
4\gamma^{2})}
\right)
\eqno (A.36)
$$
\\
{}From (A.35), (A.36) it follows

$$
L^{2}\ll 1 \  : \ \
\tilde a=\left(\frac{\pi}{2L}\right)^{2}(1-\tilde\nu)+\sqrt{1+\tilde\nu}
\left(1-\frac{\varepsilon^{2}}{(1+\tilde\nu)\tilde\nu}\left(1-\frac{2}{\pi}
\frac{\sqrt{1-\nu^{2}}}{
\tilde\nu}\cot\left(\frac{\pi}{2}\sqrt{\frac{1-\tilde\nu}{1+
\tilde\nu}}\right)\right)\right)
\eqno (A.37)
$$

\begin{figure}
\caption{The orientations of crystalline axes and the magnetic field relative
to
the film boundaries.}
\label{fig.1}
\caption{The dependence $H_{c2}(T)$ for a film from superconductor with
two-component order parameter for $\nu=\varepsilon=0.5$ and film thickness
$2d=20\xi_0$ in
 the case of speculary reflecting surfaces (1) and of diffusive electron
 scattering (2).}
\label{fig.2}
\end{figure}
\newpage

\section*{TABLES}
TABLE I.

\begin{center}
\begin{tabular}{|l|l||l|l|}
\hline
\multicolumn{2}{|c||}{Singlet} & \multicolumn{2}{c|}{Triplet} \\ \hline

$A_{1g}$&1&$A_{1u}$&$\hat{c}p_z$\\ \hline
$A_{2g}$&Im$(p_x+ip_y)^6$   & $A_{2u}$&$\hat{c} p_z$Im$(p_x+ip_y)^6 $\\ \hline
$B_{1g}$&$p_z$Im$(p_x+ip_y)^3$& $B_{1u}$&$\hat{c}$Im$(p_x+ip_y)^3$\\ \hline
$B_{2g}$&$p_z$Re$(p_x+ip_y)^3$& $B_{2u}$&$\hat{c}$Re$(p_x+ip_y)^3$\\ \hline
$E_{1g}$&$p_zp_x, p_zp_y$ &$E_{1u}$&$\hat{c}p_x,\hat{c}p_y$\\ \hline
$E_{2g}$&$2p_xp_y,p_x^2-p_y^2$&$E_{2u}$&
$\hat{c}2p_xp_yp_z,\hat{c}p_z(p_x^2-p_y^2)$\\ \hline \end{tabular}
\end{center}
\vspace{2cm}

TABLE 2.

\begin{center}
\begin{tabular}{|l|l|l|}
\hline
Representation&$\phi_0=n\pi/3$&$\phi_0=(2n+1)\pi/6$\\
\hline
$A_{1g},A_{1u}$&$\eta'(0)=0$&$\eta'(0)=0$\\
\hline
$A_{2g},A_{2u}$&$\eta(0)=0$&$\eta(0)=0$\\
\hline
$B_{1g},B_{1u}$&$\eta'(0)=0$&$\eta(0)=0$\\
\hline
$B_{2g},B_{2u}$&$\eta(0)=0$&$\eta'(0)=0$\\
\hline
$E_{1g},E_{1u}$&$\tilde{\eta}_1(0)=0$&$\tilde{\eta}_1(0)=0$\\
      &$\tilde{\eta}_2'(0)=0$&$\tilde{\eta}_2'(0)=0$\\ \hline
$E_{2g},E_{2u}$&$\tilde{\eta}_1(0)=0$&$\tilde{\eta}_1(0)=0$\\
      &$\tilde{\eta}_2'(0)=0$&$\tilde{\eta}_2'(0)=0$\\ \hline
\end{tabular}
\end{center}

\newpage
\vspace*{5cm}
\begin{picture}(110,50)
\put(52,10){\line(0,1){70}}
\put(82,10){\line(0,1){70}}
\put(67,60){\vector(0,1){20}}
\put(48,42){-d}
\put(83,42){d}
\put(110,42){x}
\put(97,52){$\vec H$}
\put(22,45){\vector(1,0){90}}
\put(92,50){\vector(1,0){15}}
\put(69,78){z}
\end{picture}
\vspace{3cm}
\begin{center}
Fig. 1
\end{center}


\begin{references}
\bibitem{bc} The boundary conditions for a tetragonal superconductor may be
derived along the same way as it has been done in Sec.II for a hexagonal
superconductor. For the given orientation with the normal to the boundary
parallel to the axis $x$ and in the case of specular reflection the boundary
conditions are independent of the particular form of the basis functions of a
representation (for the triplet pairing spins are assumed to be ordered along
$z$-axis). Then one obtains
$\eta'(0)=0$ for $A_{1g},A_{1u},B_{1g},B_{1u}$ representations, $\eta(0)=0$ for
the remaining one-dimensional representations of $D_{4h}$. For two-dimensional
representations one gets Eq.(\ref{eqno (3.6)}).
\bibitem{snos} Note that the same strict condition appears for the other
orientation when the crystalline axis z is perpendicular to the boundary and
$\bbox{H}\parallel z$.
\end{references}
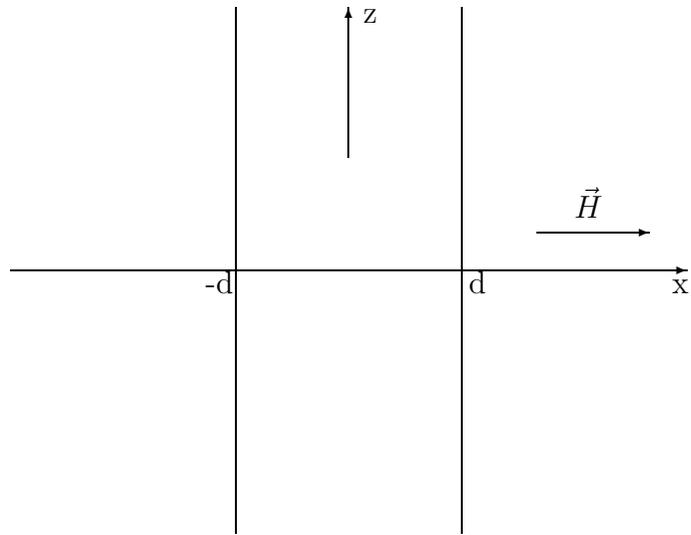
\end{document}